\documentclass[11pt]{article} 

\usepackage[ansinew]{inputenc}
\usepackage{amsmath, amssymb, graphics, amsthm}
\usepackage{epsfig}
\usepackage{color, xcolor}
\usepackage{fancyhdr} 
\usepackage{manfnt}
\usepackage[T1]{fontenc}

\usepackage[normalem]{ulem}

\definecolor{darkgreen}{rgb}{0.0, 0.4, 0.3}

\newcommand{\RB}{\mathbb R_\text{Bohr}}

\newcommand{\bra}[1]{\left\langle #1 \right|}

\newcommand{\ket}[1]{\left| #1 \right\rangle}

\newcommand{\braket}[2]{\left\langle \vphantom {#1 #2} #1 \hphantom{|} \right| \left. \vphantom {#1 #2} #2 \right\rangle}

\newcommand{\braopket}[3]{\left\langle \vphantom {#1 #2 #3} #1 \hphantom{|} \right| #2 \left| \hphantom{|} \vphantom {#1 #2 #3} #3 \right\rangle}

\makeatletter
\@addtoreset{equation}{section}
\makeatother

\newcommand{\be}{\begin{equation}}
\newcommand{\ee}{\end{equation}}
\newcommand{\ba}{\begin{eqnarray}}
\newcommand{\ea}{\end{eqnarray}}

\def\pb#1{\rlap{\lower1.5ex\hbox{$\longleftarrow$}}{#1}}
\def\dpb#1{\rlap{\lower1.5ex\hbox{$\Longleftarrow$}}{#1}}
\def\spb#1{\rlap{\lower1.0ex\hbox{$\leftarrow$}}{#1}}
\def\sdpb#1{\rlap{\lower1.0ex\hbox{$\Leftarrow$}}{#1}}

\DeclareMathOperator{\sign}{sign}

\usepackage[footnotesize,it]{caption}

\def\bq{\begin{equation}}
\def\eq{\end{equation}}
\def\d{\mathrm{d}}
\def\A{{A}}
\def\E{{E}}

\def\q{\tilde{q}}
\def\R{\tilde{R}}
\def\PR{\tilde{P}_{{R}}}
\def\tr{\mathrm{tr}}
\def\Hop{\widehat{{H}_r}}
\newcommand{\sixj}[6]{\begin{Bmatrix}  #1&#3&#5\\#2&#4&#6\end{Bmatrix}}

\newcommand{\threej}[6]{\begin{pmatrix}  #1&#3&#5\\#2&#4&#6\end{pmatrix}}
\newcommand{\makeSymbol}[1]{\mathord{\vcenter{\hbox{#1}}}}

\title{{\sf On the relation between reduced quantisation and quantum reduction for spherical symmetry in loop quantum gravity}}
\author{
{\sf N. Bodendorfer}\thanks{{\sf 
norbert.bodendorfer@fuw.edu.pl}},
{\sf A. Zipfel}\thanks{{\sf 
antonia.zipfel@fuw.edu.pl}}\\
\\
{\sf  Faculty of Physics, University of Warsaw, Pasteura 5, 02-093, Warsaw, Poland}\\
}
\date{{\small\sf \today}}

\begin{document} 

\maketitle

{\sf

\begin{abstract}

Building on a recent proposal for a quantum reduction to spherical symmetry from full loop quantum gravity, we investigate the relation between a quantisation of spherically symmetric general relativity and a reduction at the quantum level. To this end, we generalise the previously proposed quantum reduction by dropping the gauge fixing condition on the radial diffeomorphisms, thus allowing to make direct contact between previous work on reduced quantisation. A dictionary between spherically symmetric variables and observables with respect to the reduction constraints in the full theory is discussed, as well as an embedding of reduced quantum states to a sub sector of the quantum symmetry reduced full theory states. On this full theory sub sector, the quantum algebra of the mentioned observables is computed and shown to {\it qualitatively} reproduce the quantum algebra of the reduced variables in the large quantum number limit for a specific choice of regularisation. Insufficiencies in recovering the reduced algebra quantitatively from the full theory are attributed to the oversimplified full theory quantum states we use.

\end{abstract}

}

\section{Introduction}

It is generally believed that computations in full loop quantum gravity are very complicated, not because there is no concrete proposal for the dynamics, but because once a quantum state is chosen, the action of the Hamiltonian (constraint) \cite{ThiemannQSD1, GieselAQG4} or the evolution operator in spin foam models \cite{FreidelANewSpin, EngleLQGVertexWith} is very cumbersome to compute. The choice of quantum state is already a hard problem in itself, since in the usually employed spin network basis we are always considering discrete approximations to the classical continuous geometry. In the case of spherical symmetry discussed in this paper, this problem can be phrased in the following way: while the total area of a sphere $S^2$ (of symmetry) would be described by a single number in a classically reduced model, what spin network state should be used to approximate the spherically symmetric geometry of $S^2$ in the quantum reduction? A very fine state (or arbitrary superpositions thereof) would lead to an intractable problem without the use of computers, while we might risk to miss effects resulting from coarse graining when applying the fundamental dynamics to coarse states.

The probably hardest way to approach this problem is to work without any choice of (classical) gauge fixing. The problem with this approach is that all information about the geometry, or symmetry reduction, has to be encoded in the spin network state, and that the ``problem of time'', e.g. the proper interpretation of the observables of Hamiltonian general relativity, has to be solved at the quantum level. On the other hand, progress with understanding the semiclassical interpretation of quantum states has been made in the context of deparametrised (or gauge fixed) models \cite{GieselAQG4, DomagalaGravityQuantizedLoop, HusainTimeAndA}, where one classically chooses a physical coordinate system via a (partial) gauge fixing of the spatial diffeomorphism and Hamiltonian constraints. While classical deparametrisation may be criticised\footnote{In general, deparametrisation can be considered physically appropriate if the involved clocks or rods are expected to behave sufficiently classically, or that their properties are irrelevant in the situation of interest. The latter is the case for the radial gauge considered in this paper in the context of a reduction to spherical symmetry: the precise angular coordinate of points on the spatial slice is irrelevant, since one averages over angular diffeomorphisms in the reduction to spherical symmetry.
} on the ground that it suppresses quantum fluctuations of the fields that are classically solved for, it solves the ``problem of time and coordinates'', so that the main open problem in this framework is to understand the dynamics, the difference between classical and quantum reduction, and the choice of coarse or fine states. 

Recently, a proposal for symmetry reductions in the context of quantising a classically gauge fixed version of general relativity was given for both for spherical symmetry and Bianchi I models \cite{BIII, BLSI}.
The main conceptual input that allowed for this development was to realise that on the reduced phase space, obtained from gauge fixing, the action of certain functions, which vanish in the symmetry reduced sector, is that of a certain set of spatial diffeomorphisms. The reduction at the quantum level can then be implemented using well-known techniques \cite{AshtekarQuantizationOfDiffeomorphism} for implementing the spatial diffeomorphism constraint in loop quantum gravity.

While in \cite{BIII} also the mini-superspace dynamics could be extracted from the full theory, 
 the problem is harder for spherical symmetry
 in the context of \cite{BLSI}, and it is unclear whether an exact equivalence to a reduced quantisation can even be achieved. We will take fist steps in investigating this problem in the present paper, using as an embedding of reduced theory quantum states into the reduced sector of the full theory the proposal of \cite{BLSI}, which uses crucial properties of the $2+1$-dimensional volume operator \cite{ThiemannQSD4}. We will define the analogues of the operators occurring in reduced quantisations as operators on the reduced sector of the full quantum theory and compute their algebra. It will turn out that the quantum algebra in the reduced models can be reproduced {\it qualitatively} in the large quantum number limit for a certain choice of regularisation, that is up to numerical factors. The current construction however has some insufficiencies as will be discussed later, and more detailed studies are needed.\\

This paper is organised as follows: we start in section \ref{sec:GeneralStrategy} by describing the general strategy. We continue in section \ref{sec:Classical} by defining our quantisation variables in a gauge fixed setting and by relating the reduced and full theory phase spaces. In section \ref{sec:QuantumTheory}, we quantise the classical theory for both the reduced and full theory case. The quantum algebra's of the reduced and full theories are compared in section \ref{sec:Comparison}. Concluding remarks follow in section \ref{sec:Conclusion}.

\section{General strategy} \label{sec:GeneralStrategy}

\subsection{Main idea}

A key role in the procedure of canonical quantisation is played by a certain Poisson sub-algebra of all phase space functions which we want to represent as operators on a Hilbert space. In particular, this algebra, along with the Hamiltonian, determines the quantum dynamics. A good starting point for comparing the dynamics of the classically or quantum reduced theories is thus to study the quantum algebra of quantum reduced operators from a full theory perspective.

Clearly, if a symmetry reduction is performed at the classical level, the available algebra of functions is smaller than in the case of the full theory. If one wants to compare the algebraic structure of the full and the symmetry reduced theory, one therefore has to express functions on the symmetry reduced\footnote{It should be kept in mind that the ``symmetry reduced'' phase space, which is obtained after considering only spherically symmetric geometries, is different from the ``reduced'' phase space, which denotes the phase space obtained after solving the constraints.} phase space by a certain choice of full phase space functions, which, when evaluated on the symmetry reduced sector of the full phase space, agree with those of the reduced theory. 

Classically, this agreement is {\it by construction}. After quantisation however, the situation is less clear. In particular, given a certain phase space function in the reduced theory, there generally exist multiple ways of writing it on the full phase space. An example that will be relevant later is that one can extract the radius of a sphere from both its surface area and its circumference. After quantisation however, the operators extracting the radius from the circumference or the area might have different properties, as we will see later in section \ref{sec:FullSpectralProperties}.

If an isomorphism between those algebras could be established, (which has been possible in this paper only partially), one has already gone a long way in showing an equivalence between the classically and quantum reduced theories. In particular, one can then argue that following the classical reduction procedure, one should take the full classical Hamiltonian and write it as a sum of the classically reduced Hamiltonian plus ``inhomogeneous'' correction terms, whose (to be properly defined) quantisation should vanish when acting on symmetric quantum states. Then, when expressing the classically reduced variables in the classically reduced Hamiltonian by full theory operators, the action of this full theory Hamiltonian would be the same as the action of the reduced Hamiltonian. In general however, one expects that such an exact matching is not directly possible as the quantum reduced theory generically captures more (quantum) degrees of freedom than the classically reduced one. Thus, some coarse graining of the quantum reduced theory is needed in order to match the classically reduced theory. This point of view is supported by the findings in this paper.

\subsection{Technical realisation}

The techniques used in \cite{BLSI} for the quantum symmetry reduction to spherical symmetry rely on a classical choice of gauge fixing for the spatial diffeomorphism constraint adapted to spherical symmetry. At the Hamiltonian level, introducing a gauge fixing corresponds to adding constraints to the theory which are second class with respect to the generator of gauge transformations, in this case the spatial diffeomorphism constraint. Solving the second class pair of constraints, or equivalently using the Dirac bracket, leads to the reduced phase space of the theory. The main observation of \cite{BLSI} was that one can construct suitable connection variables which parametrise the reduced phase space, on which constraints compatible with a reduction to spherical symmetry act via a sub class of spatial diffeomorphisms. The spatial diffeomorphism constraint, which can be solved at the quantum level via the techniques of \cite{AshtekarQuantizationOfDiffeomorphism}, thus reappears in the theory, however enforcing a reduction of physical degrees of freedom, as opposed to removing gauge redundancy. Observables with respect to the reduction constraints can thus easily be identified, e.g. as the area of spheres of constant radius.
A comparison between the action of operators in the classically reduced theory and their counterparts in the reduced sector of the full quantum theory can thus be explicitly made.

Care has to be taken however in the choice of variables, i.e. one needs to construct variables in the classically reduced theory which are as close as possible to those in the full theory. While there is a great deal of freedom in choosing quantisation variables in the classically reduced theory, this is not the case in the full theory as we will see. There, one needs to pay special attention to the density weight of the connection, which leaves only one possibility to define properly transforming variables that are structurally similar to the Ashtekar-Barbero variables of full loop quantum gravity.

\section{Classical preparations}

\label{sec:Classical}

\subsection{Gauge fixing and reduced phase space}

In this section, we will closely follow the ideas of \cite{BLSI, DuchObservablesForGeneral, BLSII}. The main difference will be that we do not gauge fix all of the the radial diffeomorphisms, i.e. the metric component $q_{rr}$ will only be demanded to be angle-independent.

We start with the ADM phase space \cite{ArnowittTheDynamicsOf}, coordinatised by the spatial metric $q_{ab}$ and its conjugate momentum $P^{ab} = \frac{\sqrt{\det{q}}}{2} \left(K^{ab} - q^{ab} K \right)$, where $K_{ab}$ is the extrinsic curvature. The non-vanishing Poisson brackets are given by $\left\{q_{ab}(x), P^{cd}(y) \right\} = \delta_{(a}^c \delta_{b)}^d \delta^{(3)}(x,y)$ and we have chosen units such that $\kappa = 8 \pi G = 1$. We denote the spatial slice by $\Sigma \ni x,y$. The spatial diffeomorphism constraint $C_a$ smeared with the shift vector $N^a$ is given by $C_a[N^a] = -2 \int_\Sigma d^3x N^a \nabla_b P^b {}_a$.

We now introduce the gauge fixing $q_{r A} = 0$ for the angular ($N^A$) components of the spatial diffeomorphism constraint\footnote{As explained in detail in \cite{DuchObservablesForGeneral, BLSII}, the coordinate system in which this gauge condition is stated refers to ``adapted coordinates'' of a fiducial metric $\check q$. This technical point is not of importance in this paper, and we will not comment on it further.}. The metric is thus constrained to have the form
\be \left( \begin{array}{ccc}
q_{rr} & 0 & ~0 \vspace{1mm} \\
0 &   & \vspace{-1mm} \\
 &  \hspace{3mm}  q_{AB}  \hspace{-5mm} & \vspace{-1mm}  \\ 
0 &  &  \end{array} \right) \text{.}\ee
As canonical variables on the reduced phase space, we choose the pairs 
\be
	\left\{q_{rr}(r,\theta), P^{rr}(r',\theta') \right\} = \delta(r,r') \delta^{(2)}(\theta,\theta')~~ \text{and} ~~ \left\{q_{AB}(r,\theta), P^{CD}(r',\theta') \right\} = \delta_{(A}^C \delta_{B)}^D \delta(r,r') \delta^{(2)}(\theta,\theta') \text{.} 
\ee
Following \cite{DuchObservablesForGeneral}, we denote by $\theta$ collectively the two angular spatial coordinates perpendicular to the radial direction. The corresponding tensor indices are denoted by $A,B,\ldots$.
Next, we have to solve $C_A[N^A] = 0$ for $P^{rA}$. Since $C_A$ is linear in $P^{ab}$, this can be done conveniently by solving the equation
\be
	\left. \frac{\delta}{\delta P^{rB}(x)} \left( P^{rA}[\lambda_{rA}] + C_A \left[N^A\right]\right)  \right|_{q_{\check r \check A}=0} = 0  \label{eq:PrAReduced}
\ee
for $N^A$ as a function of the arbitrary smearing functions $\lambda_A$. The virtue of this equation is that $P^{rA}[\lambda_{rA}] + C_A \left[N^A[\lambda_{rB}]\right]$ does not depend on $P^{rA}$ any more, while it has been modified only by terms proportional to $C_A$. We note again that this follows form linearity of $C_A$ in $P^{ab}$ and that equation \eqref{eq:PrAReduced} can be considered as the first contribution form a gauge unfixing projector \cite{MitraGaugeInvariantReformulationAnomalous, VytheeswaranGaugeUnfixingIn}, computing a gauge (generated by $q_{\check r \check A} = 0$) invariant extension of $P^{rA}$ by modifying it with terms proportional to $C_A$. This however is exactly equivalent to solving $C_A = 0$ for $P^{rA}$. 
 
The solution to \eqref{eq:PrAReduced} can be inferred directly from the more general results of \cite{DuchObservablesForGeneral}, which were also used in \cite{BLSI, BLSII}. In particular, solving \eqref{eq:PrAReduced} is equivalent to the analogous problem in \cite{BLSI, BLSII} for $\lambda_{rr} = 0$, i.e. not also solving $C_r=0$ for $P^{rr}$. We find, up to terms proportional to $\partial_A q_{rr}$ which we will drop due to the later introduction of the gauge fixing condition \eqref{eq:LambdaAngleConstraint},
\be
	N^A(r, \theta) = \int_0^r dr' q^{AB}(r', \theta) \lambda_{rB}(r', \theta)
\ee
and conclude that sampling over all possible $\lambda_{rB}$, we can generate arbitrary $N^A$. Our strategy to reduce to spherical symmetry at the quantum level will thus be to impose 
\be
	\left. P^{rA}[\lambda_{rA}] + C_A \left[N^A[\lambda_{rB}]\right] \right|_{q_{rA} = 0} =0
\ee
as an operator equation for all $\lambda_{rA}$. Since the previous expression however does not depend on $P^{rA}$ any more, this is equivalent to imposing 
\be
	\left. C_A \left[N^A\right]  \right|_{q_{rA} = 0, ~ P^{rA} = 0}	=0\label{eq:AngularDiffReduced}
\ee
for arbitrary $N^A$. Let us make a few remarks about our strategy. As advocated in \cite{BIII, BLSI}, we want to impose classical conditions implied by spherical symmetry at the quantum level. We certainly cannot implement all such conditions as strong operator equations, since they include second class pairs. Thus, a certain choice of a commuting subset is always necessary.  Another question that one might ask is about whether the classical analogues of the conditions imposed reduce the classical theory precisely to spherical symmetry. While this might be desirable, it is logically not necessary, might not lead to the easiest implementation of spherical symmetry, and might even be impossible to achieve in general. In our case, imposing $P^{rA} =0$ is certainly consistent with spherical symmetry as discussed in \cite{KucharGeometrodynamicsOfThe} and it is implementable via standard techniques. Moreover, after implementing $P^{rA} = 0$, the theory will already be reduced further than another plausible definition of quantum spherical symmetry given in \cite{BLSI}, based on group averaging over rigid SO$(3)$ rotations around the central observer. The procedure will thus be justified a posteriori by its result, the possibility to implement a plausible definition of quantum spherical symmetry.

Let us now compute equation \eqref{eq:AngularDiffReduced} explicitly:
\begin{align}
	\left. C_A \left[N^A\right]  \right|_{q_{rA} = 0,~ P^{rA} = 0} &= - 2  \int_\Sigma dr \, d^2 \theta \, N^A \left( \partial_B P^B {}_A - \stackrel{(3)}{\Gamma _{BA}^C} P^B {}_C - \stackrel{(3)}{\Gamma_{rA}^r} P^r {}_r \right) \nonumber \\
	&= -2 \int_\Sigma dr \, d^2 \theta \, N^A \left( \stackrel{(2)}{\nabla}_B P^{B} {}_A  -\frac{1}{2} P^{rr} \partial_A q_{rr} \right) \nonumber \\
	&= \int_\Sigma dr \, d^2 \theta \, \left( P^{AB} \mathcal L_{\vec N} q_{AB} + P^{rr} \mathcal L_{\vec N} q_{rr} \right) \text{,}
\end{align}
where $\stackrel{(3)}{\Gamma}$ is the three-dimensional Christoffel symbol derived from $q_{ab}$, $\stackrel{(2)}{\nabla}_A$ is the covariant derivative compatible with $q_{AB}$, acting on angular indices $A,B$, and $\mathcal L_{\vec N}$ is the Lie derivative with respect to $\vec N = N^A \partial_A$. The usual boundary term giving the ADM momentum drops because the partial integration is only in angular directions. We conclude that \eqref{eq:AngularDiffReduced} acts on the reduced phase space via spatial diffeomorphisms preserving spheres of constant $r$, where $q_{rr}$ transforms as a scalar, $P^{rr}$ as a density, $q_{AB}$ as a covariant symmetric $2$-tensor, and $P^{AB}$ as densitised contravariant symmetric $2$-tensor.

Next, let us express $C_r[N^r]$ in terms of the reduced phase space coordinates, separating away terms proportional to $P^{rA}$: 

\begin{align}
	\left. C_r \left[N^r\right]  \right|_{q_{rA} = 0} &= - 2  \int_\Sigma dr \, d^2 \theta \,  N^r \left( \partial_r P^r {}_r - \stackrel{(3)}{\Gamma _{Br}^C} P^B {}_C - \stackrel{(3)}{\Gamma_{rr}^r} P^r {}_r \right) + \mathcal{O}\left(P^{rA} \partial_B N^r \right) \nonumber \\
	&= -2 \int_\Sigma dr \, d^2 \theta \,  N^r \left( \partial_r \left( P^{rr} q_{rr} \right) - \frac{1}{2} P^{AB} \partial_r q_{AB} - \frac{1}{2}P^{rr} \partial_r q_{rr} \right) + \mathcal{O}\left(P^{rA}\partial_B N^r\right)  \nonumber \\
	&= \int_\Sigma dr \, d^2 \theta \,  \left( P^{AB} \mathcal L_{\vec N} q_{AB} + P^{rr} \mathcal L_{\vec N} q_{rr} \right) + \mathcal{O}\left(P^{rA} \partial_B N^r\right)   \text{,} \label{eq:RadialDiffeoConstraint}
\end{align}
where in the last line we have included the boundary term leading to the ADM-momentum. 
We see that $C_r$ again acts via spatial diffeomorphisms, however in radial direction, and up to the terms proportional to $P^{rA} \partial_B N^r$. It should be noted that $q_{rr}$ transforms as a density of weight $+2$ under radial diffeomorphisms, whereas $q_{AB}$ transforms as a scalar. Consequently, $P^{rr}$ transforms as a density of weight $-1$, and $P^{AB}$ as a density of weight $+1$. We will not detail the terms proportional to $P^{rA}$, since they will drop later on due to the further gauge fixing \eqref{eq:LambdaAngleConstraint} which implies $\partial_B N^r = 0$ for the remaining radial diffeomorphisms.

It will be useful in the following to change variables from $q_{rr}$ to $\Lambda$ with $\Lambda^2 = q_{rr}$. We denote the conjugate momentum of $\Lambda$ by $P_\Lambda := 2 \sqrt{q_{rr}} P^{rr}$, so that the new Poisson brackets read 
\be 
	\left\{\Lambda(r,\theta), P_\Lambda(r',\theta') \right\} = \delta(r,r') \delta^{(2)}(\theta,\theta')~~ \text{and} ~~ \left\{q_{AB}(r,\theta), P^{CD}(r',\theta') \right\} = \delta_{(A}^C \delta_{B)}^D \delta(r,r') \delta^{(2)}(\theta,\theta') \text{.} 
\ee
These Poisson brackets in particular agree with the Dirac brackets that one would obtain from implementing the gauge fixing (see \cite{BLSI, BLSII} for details).
Our $\Lambda$ now corresponds to the $\Lambda$ of \cite{KucharGeometrodynamicsOfThe}, where the spherically symmetric ansatz $ds^2 = \Lambda^2 (r,t) dr^2 + R^2(r,t) d \Omega^2$ is made for the spatial line element. In these variables, the Hamiltonian constraint reads 
\begin{align} \nonumber
	H[N]&= \int d^3x \, N \left( \frac{2}{\sqrt{q}} \left( P^{ab} P_{ab} - \frac{1}{2} P^2 \right) - \frac{\sqrt q}{2} \stackrel{(3)} R\right)\\
	 &= \int_\Sigma dr \, d \theta \, N \Biggr( \frac{1}{\Lambda \sqrt{\det{q_{AB}}}} \left(  \frac{P_\Lambda^2 \Lambda^2}{4} - P_\Lambda \, \Lambda \,P^{A} {}_A + 2 P^{AB} P_{AB} -    P^{A} {}_A P^{B} {}_B \right) \nonumber \\
	&~~~~~~~~~~~~~~~~~~~~~ - \frac{ \Lambda \sqrt{\det{q_{AB}}}}{2}  \biggr(\stackrel{(2)}{R} - \frac{1}{4 \Lambda^2}  (\partial_r q_{AB}) (\partial_r q_{CD}) \left( q^{AB} q^{CD} + q^{AC} q^{BD} \right) \nonumber ~~~~~~~~~~\\
	& ~~~~~~~~~~~~~~~~~~~~~~~~~~~~~~~~~~~~~~~~~~~- \Lambda^{-1}\partial_r (q^{AB} \Lambda^{-1} \partial_r q_{AB}) \biggr) \Biggr) + \mathcal{O}(P^{rA}) \text{.} \label{eq:HamiltonianConstraint}
\end{align}
We now perform another change of variables, which will ultimately lead to connection variables transforming with the correct density weights in the next section. At this point however, its purpose is less transparent. We define
\be
	\tilde q_{AB} := \Lambda^2 q_{AB}, ~~ \tilde P^{AB} := \Lambda^{-2} P^{AB}, ~~ \tilde P_\Lambda = P_\Lambda - \frac{2 P^{AB} q_{AB}}{\Lambda}
\ee
so that the new non-vanishing Poisson brackets read
\be
	\left\{\Lambda(r, \theta), \tilde P_\Lambda (r', \theta') \right\} = \delta(r,r')\delta^{(2)}(\theta,\theta')~~ \text{and} ~~ \left\{\tilde q_{AB}(r,\theta), \tilde P^{CD}(r',\theta') \right\} = \delta_{(A}^C \delta_{B)}^D \delta(r,r') \delta^{(2)}(\theta,\theta')  \text{.} 
\ee

It will be convenient later to introduce a partial gauge fixing of the radial diffeomorphisms such that only angle independent shift vectors $N^r = N^r(r)$ generate gauge transformations. To this end, we introduce the additional constraints 
\be
	\Lambda(r,\theta) - \Lambda(r,\theta') = 0 ~~ \forall ~ r,\theta, \theta' \text{,} \label{eq:LambdaAngleConstraint}
\ee 
which enforce that points with the same radial coordinate $r$ have the same proper distance $\int_0^r | \Lambda(r)| dr$ from the origin $\sigma_0$. $C_r$ smeared with a shift vector with non-trivial angle dependence is thus automatically second class with \eqref{eq:LambdaAngleConstraint}, whereas $C_r$ smeared with a shift vector with only radial dependence remains first class. We note that \eqref{eq:LambdaAngleConstraint} is also first class with angular diffeomorphisms, since $\Lambda$ transforms as a scalar thereunder.

We now in principle have to modify the Hamiltonian constraint to preserve this new gauge fixing by adding $C_r$ smeared with a shift vector tailored to stabilise \eqref{eq:LambdaAngleConstraint}. For the purpose of this paper, we however get around this by the following argument.
The quantum states that we will be interested in later in this paper have the property that the Hamiltonian constraint acts non-trivially only at a single point on each $S^2_r$, so that it is sufficient to consider lapse functions with only radial dependence. Computing the action of the Hamiltonian constraint on $\Lambda$, we find
\be
	\left\{ \Lambda, H[N]  \right\} = \frac{N \Lambda \tilde P_\Lambda}{2 \sqrt{\det{\tilde q_{AB}}}}:= \dot \Lambda 
\ee
Inspecting $\dot \Lambda$, we see that if $P^{rA} = 0$ is imposed, ensuring angular diffeomorphism invariance of $\tilde P^{AB}$, $\tilde q_{AB}$, $\Lambda$, as well as $\tilde P_\Lambda$, and the condition $N = N(r)$ holds, \eqref{eq:LambdaAngleConstraint} is already preserved without adding terms proportional to $C_r$ to $H$. 
If we were now to admit angle-dependent lapse functions, we would have to stabilise the Hamiltonian by adding the angle-dependent radial diffeomorphisms, however as said above, angle-independent lapse functions will be sufficient in the quantum theory due to our choice of quantum states to be discussed in section \ref{sec:FullQuantisation}. The shift vector generated by stabilising \eqref{eq:LambdaAngleConstraint} would simply vanish when acting on the reduced set of spherically symmetric full theory states. Intuitively, they would need to be used to adjust the spheres $S^2_r$ such that all their points have the same proper distance from $\sigma_0$, which is however already the case in spherical symmetry.

According to the procedure of gauge unfixing \cite{MitraGaugeInvariantReformulationAnomalous, AnishettyGaugeInvarianceIn}, we can now view the angle-dependent radial diffeomorphisms as gauge fixing conditions of \eqref{eq:LambdaAngleConstraint} and drop them. In turn, we can gauge fix \eqref{eq:LambdaAngleConstraint} to simplify the theory further.
We demand that
\be
	\frac{\tilde P_\Lambda(r,\theta)}{\sqrt{\det{\tilde q_{AB}(r,\theta)}}} - \frac{\tilde P_\Lambda(r,\theta')}{\sqrt{\det{\tilde q_{AB}(r,\theta')}}}  = 0 ~~ \forall ~ r,\theta, \theta' \text{.} \label{eq:PLambdaAngleConstraint}
\ee 
The constraints \eqref{eq:LambdaAngleConstraint} and \eqref{eq:PLambdaAngleConstraint} can be solved simply by restricting the phase space coordinates $\left(\Lambda(r,\theta), P_\Lambda(r,\theta) \right)$ to $\left(\Lambda(r), P_\Lambda(r) \right)$ with the non-vanishing Poisson brackets
\be
	\left\{\Lambda(r), \tilde P_\Lambda (r') \right\} = \delta(r,r')~~ \text{and} ~~ \left\{\tilde q_{AB}(r,\theta), \tilde P^{CD}(r',\theta') \right\} = \delta_{(A}^C \delta_{B)}^D \delta(r,r') \delta^{(2)}(\theta,\theta') \text{.} 
\ee
The relation to the original phase space coordinates is 
\be
	\Lambda(r,\theta) = \Lambda(r) ~~ \text{and} ~~ \tilde P_\Lambda(r, \theta) = \tilde P_\Lambda(r) \frac{\sqrt{\det{\tilde q_{AB}(r,\theta)}}}{4 \pi \Lambda^2(r) R(r)^2}
\ee
with
\be
	4 \pi \Lambda(r)^2 R(r)^2 := \int_{S^2_r} d^2 \theta \sqrt{\det{\tilde q_{AB}(r,\theta)}} \text{.}
\ee

At this point, we are thus left with the the phase space coordinatised by $q_{AB}(r, \theta), P^{CD}(r, \theta)$, and $\Lambda(r), P_\Lambda(r)$. It is subject to the Hamiltonian constraint \eqref{eq:HamiltonianConstraint}, corrected with radial diffeomorphism constraints in case of lapse functions with non-trivial angular dependence, and the radial part of the spatial diffeomorphism constraint \eqref{eq:RadialDiffeoConstraint} smeared with angle-independent shift vectors. 
The precise form of the Hamiltonian constraint in the new variables is not of importance for this paper and we will not spell it out. What is important is the following form of the spatial diffeomorphism constraint in the new variables with the conditions $q_{rB}=P^{rB}=0$ imposed:
\begin{align}
	\left. C_A[N^A] \right|_{q_{rB}=P^{rB}=0} &= \int_\Sigma dr \, d^2 \theta \, \tilde P^{AB} (r,\theta )\mathcal L_{\vec N} \tilde q_{AB} (r,\theta)	\\
	\left. C_r[N^r]  \right|_{q_{rB}=P^{rB}=0}& = \int_\Sigma dr \, d^2 \theta \, \tilde P^{AB} (r,\theta )\mathcal L_{\vec N} \tilde q_{AB} (r,\theta) + \int_{\mathbb R^+_0} dr \tilde P_\Lambda (r) \mathcal L_{\vec N} \Lambda (r)
\end{align}
We note that $\tilde q_{AB}$ now behaves as a density of weight $+2$ under radial diffeomorphisms, and $\tilde P^{AB}$ as a density of weight $-1$. $\Lambda$ has radial density weight $+1$, and $\tilde P_\Lambda$ accordingly $0$. The other (angular) density weights are as before.

This phase space still corresponds to full general relativity, restricted to spacetimes allowing for the gauge $q_{rA} = 0$ and $\partial_A q_{rr}=0$. On top, we have identified the action of $P^{rA}$ on the reduced phase space as the action of spatial diffeomorphisms preserving the spheres $S^2_r$ of constant (proper) radius, i.e. geodesic distance from $\sigma_0$. More explicitly in the new variables, we have
\be
	P^{rA} = 0 \Leftrightarrow C_A[N^A] |_{q_{rB}=P^{rB}=0}  = 0 \text{.}
\ee
While we will impose $P^{rA}=0$ in the quantum theory to reduce to spherical symmetry, this is not necessary from a conceptual point of view, we could also choose to look at full general relativity in this gauge.

\subsection{Full theory canonical variables}
\label{sec:FullTheoryCanonicalVariables}

In order to apply the quantisation methods of loop quantum gravity to our theory, we need to coordinatise the above phase space by connection type variables and their momenta. As was already noted in \cite{BLSI}, $\tilde q_{AB}$ and $\tilde P^{AB}$ are naturally suited to be replaced by SU$(2)$ connection variables familiar from three-dimensional Euclidean gravity, modified by a free Barbero-Immirzi-type parameter. In the context of LQG, these variables were used for example in \cite{ThiemannQSD4, Wisniewski2+1GeneralRelativity}, and they are precisely a restriction to $2+1$ dimensions of the dimension-independent connection variables defined in \cite{BTTI}.

We start by enlarging the phase space to admit a local SU$(2)$ gauge symmetry. The non-vanishing Poisson-brackets between the new phase space variables are 
\be
	\left\{\tilde K_A^i (r, \theta), \tilde E^B_j (r', \theta') \right\} = \delta(r,r')\delta^{(2)}(\theta,\theta') \delta_A^B \delta^i_j \text{,}
\ee 
where $i,j$ are SU$(2)$ Lie algebra indices, and the relation to the former phase space variables is 
\be
	(\det \tilde q) \tilde q^{AB} = \tilde E^{Ai} \tilde E^{B}_i, ~~~ \sqrt{\det \tilde q} \stackrel{(\tilde P)}{K}_{AC} \tilde q^{BC}  = \tilde K_{Ai} \tilde E^{Bi}\text{,}
\ee
with
\be
	\tilde P^{AB} = \frac{\sqrt{\det \tilde q}}{2} \left( \stackrel{(\tilde P)}{K} {}^{AB} - \tilde q^{AB}  \stackrel{(\tilde P)}{K}_{CD} \tilde q^{CD} \right) \text{.} \label{eq:DefOfKP}
\ee
All raising and lowering of the $A,B$ indices is now performed with the metric $\tilde q_{AB}$. We note that $\stackrel{(\tilde P)}{K}_{AB}  \neq K_{AB}$, that is $\stackrel{(\tilde P)}{K}_{AB}$ is not the extrinsic curvature, but an object that is obtained from $\tilde P^{AB}$ seeing it as the momentum conjugate to the ``two-metric'' $\tilde q_{AB}$, and then extracting the extrinsic curvature from it via the usual definition \eqref{eq:DefOfKP} stemming from the ADM formulation of three-dimensional general relativity. Our new phase space is now subject to the Gau{\ss} law 
\be
	G_{ij} := \tilde E^{A}_{[i} \tilde K_{A|j]} = 0 \text{,}
\ee
which deletes the additional degrees of freedom that we have introduced. In particular, $\tilde q_{AB}$ and $\tilde P^{AB}$ are observables with respect to $G_{ij}$. 

We are now in the position to perform a canonical transformation to connection variables. We introduce the Peldan type hybrid spin connection $\tilde \Gamma_{Aij}$, defined via \cite{PeldanActionsForGravity}
\be
	\partial_A \tilde E^{Bi} + \stackrel{(2)}{\tilde \Gamma} \hspace{-1.5mm} {}_{AC}^B \tilde E^{Ci} -\stackrel{(2)}{\tilde \Gamma}  \hspace{-1.5mm}{}_{AC}^C \tilde E^{Bi} + \tilde \Gamma_{A} {}^i {}_j \tilde E^{Bj} = 0 \text{.}
\ee
From it, we construct the connection $A^i_{A} = - \frac{1}{2} \epsilon^{ijk} \tilde \Gamma_{Ajk} + \beta \tilde K_A^i =: \Gamma_A^i + \beta \tilde K_A^i$, where $\beta \in \mathbb R \backslash \{0\}$ is a free parameter. 
The covariant derivative with respect to $A_A^i$ can be written as $D_A v^{i} = \partial_A v^{i} + \epsilon^{ijk} A_{Aj} \tilde v_k$.
In order to simplify notation in the following, we define $E^{Ai} =\tilde E^{Ai} / \beta$, keeping in mind that $E^{Ai}$ does not correspond to an ``un-tilded'' $\tilde E^{Ai}$, i.e. rescaled with $\Lambda$, but $\tilde E^{Ai}$ rescaled with the free parameter $\beta$, and similar for $\tilde A_{A}^i$. The new Poisson brackets read
\be
	\left\{A_{A}^i(r, \theta),  E^B_j(r', \theta') \right\} = \delta(r,r')\delta^{(2)}(\theta,\theta') \delta_A^B \delta^i_j  \label{eq:PoissonBracketAE}
\ee
and the spatial diffeomorphism and Gau{\ss} constraints are given by
\begin{align}
	\left. C_A[N^A] \right|_{q_{rB}=P^{rB}=0} &= \int_\Sigma dr \, d^2 \theta \, E^{A}_i	 (r,\theta )\mathcal L_{\vec N} A_A^i (r,\theta) \label{eq:DiffConstrAngConnection}	\\
	\left. C_r[N^r]\right|_{q_{rB}=P^{rB}=0} & = \int_\Sigma dr \, d^2 \theta \, E^{A}_i	 (r,\theta )\mathcal L_{\vec N} A_A^i (r,\theta) + \int_{\mathbb R^+_0} dr \tilde P_\Lambda (r) \mathcal L_{\vec N} \Lambda (r)	\label{eq:DiffConstrRadConnection}	\\
	G^{i}[\lambda_{i}] &:= -G_{ij}[\epsilon^{ijk} \lambda_{k}] = \int_\Sigma dr \, d^2 \theta \, \lambda_i D_A E^{Ai} \label{eq:GaussLaw} \text{.}
\end{align}
Again, the precise form of the Hamiltonian constraint does not matter for this paper and we will not present it here. 
In \eqref{eq:DiffConstrAngConnection}, $A_A^i$ is a one-form with respect to the Lie derivative in angular direction. Consequently, $E^A_i$ is a densitised vector. 
On the other hand, in \eqref{eq:DiffConstrRadConnection}, $A_A^i$ is a scalar with respect to the Lie derivative in radial direction, and $\Lambda$ a density of weight $+1$. Consequently, $E^{A}_i$ has density weight $+1$ and $\tilde P_\Lambda$ has density weight $0$. The transformation property under finite diffeomorphisms is analogous, since it is derived from the action of $C_r$ and $C_A$. 

$E^A_i$ can thus be naturally smeared over a two-surface which is foliated by radial geodesics, while $A_A^i$ should be integrated along paths of constant $r$. This will allow us to construct spin networks with all edges embedded in a respective $S^2_r$, while the fluxes act non-trivially on corresponding holonomies.
These ``angular'' spin networks, after having been diffeomorphism averaged over the $S^2_r$ by the virtue of $P^{rA} = 0$, can then be averaged over radial diffeomorphisms generated by \eqref{eq:DiffConstrRadConnection}, under which they transform as scalars. The details will be discussed in section \ref{sec:FullQuantisation}.

\subsection{Different choices of spherically symmetric variables}

\label{sec:ChoicesOfReducedVariables}

The central aim of this paper is to compare a reduced quantisation with a quantum reduction for the case of spherical symmetry in general relativity. To this end, we also need to study the reduction to spherical symmetry at the classical level, which will be the aim of this section. We will discuss several possible choices of variables and compare their details at the end of this section.

The phase space coordinates $\Lambda(r)$ and $\tilde P_\Lambda(r)$ can already be considered as classically reduced variables, although they were derived via a gauge fixing that doesn't remove physical degrees of freedom. Let us now reduce also the other canonical pair. This can be done conveniently at the level of $q_{AB}$, $P^{AB}$ by the ansatz
\be
	q_{AB} = R^2(r) \Omega_{AB} \text{,}
\ee
where $\Omega_{AB}$ is the standard spherically symmetric metric on $S^2$. At the level of the symplectic potential, it follows that
\be
	\int_\Sigma dr \, d^2 \theta \, P^{AB} \delta q_{AB} = \int_\Sigma dr \, d^2 \theta \, P^{AB} \Omega_{AB} \delta R^2 = \int_\Sigma dr \, d^2 \theta \, \frac{2 P^{AB} q_{AB}}{R} \delta R =: \int_{\mathbb R^+_0} dr \,  P_R \delta R, {\text{,}}  \label{eq:DerivationPR}
\ee
so that $P_R (r) = \int_{S^2} d^2 \theta \, \frac{2 P^{AB} q_{AB}}{R} (r,\theta) $ and $4 \pi R^2(r) = \int_{S^2} d \theta \sqrt{\det q_{AB}}$. In these variables, the constraints reduce to\footnote{Here, $P_\Lambda$ always refers to $P_\Lambda(r) = \int_{S^2} d \theta P_\Lambda(r,\theta)$. We also abbreviated $R' := \partial_r R$ etc.}
\begin{align}
	H[N] &= 8 \pi \int_{\mathbb R^+_0} dr \,  N \left( \frac{1}{2} \frac{P_\Lambda^2 \Lambda}{(8 \pi)^2 R^2}- \frac{P_{\Lambda} P_R}{(8 \pi)^2 R} - \frac{\Lambda(r)}{2} + \frac{R'^2}{2 \Lambda} + \frac{R R''}{\Lambda} - \frac{R R' \Lambda'}{\Lambda^2}\right)				\\
	C_r[N^r] & = \int_{\mathbb R^+_0} dr  \left(  P_r \mathcal L_{\vec N}  R+  P_\Lambda \mathcal L_{\vec N} \Lambda \right)	\text{.}
\end{align}
The density weight of $\Lambda$ and $P_R$ is $+1$, the density weight of $P_\Lambda$ and $R$ is zero and $\vec N = N^r \partial_r$.

Reduced variables corresponding to our choice of full theory variables from the previous section can be obtained as
\be
	\int_{\mathbb R^+_0} dr \left( P_R \delta R + P_\Lambda \delta \Lambda \right) = \int_{\mathbb R^+_0} dr \left( \frac{P_R}{\Lambda} \delta (\Lambda R) + \left( P_\Lambda - \frac{R P_R}{\Lambda} \right) \delta \Lambda \right) =: \int_{\mathbb R^+_0} dr \left( \tilde P_R \delta \tilde R + \tilde P_\Lambda \delta \Lambda \right) \text{.}
\ee 
Their exact relation to the full theory variables will be discussed in the next subsection. We have
\begin{align}
	H[N] &= 8 \pi \int_{\mathbb R^+_0} dr \, N \left( \frac{\Lambda^3 \tilde P_\Lambda^2}{2 (8 \pi)^2 \tilde R^2} - \frac{\Lambda \tilde P_R^2}{2 (8\pi)^2} - \frac{\Lambda}{2}  + \frac{7 \tilde R^2 ( \Lambda')^2}{2\Lambda^5} + \frac{(\tilde R ')^2}{2\Lambda^3} + \frac{\tilde R \tilde R''}{\Lambda^3} - \frac{4 \tilde R \tilde R' \Lambda'}{\Lambda^4} - \frac{\tilde R^2 \Lambda''}{\Lambda^4}\right)				\\
	C_r[N^r] & = \int_{\mathbb R^+_0} dr \left( \tilde P_r \mathcal L_{\vec N} \tilde R+ \tilde P_\Lambda \mathcal L_{\vec N} \Lambda \right) \text{,}
\end{align}
with $\tilde P_R$ now transforming as a scalar under radial diffeomorphisms, while $\tilde R$ transforms as a density. 

Another popular choice of connection variables can be derived by reducing a connection formulation of general relativity in terms of Ashtekar-Barbero variables, see e.g. \cite{BojowaldSymmetryReductionFor}. One has
\begin{align}
	\int_{\mathbb R^+_0} dr \left( P_R \delta R + P_\Lambda \delta \Lambda \right) &= \int_{\mathbb R^+_0} dr \left( \left( P_R - \frac{P_\Lambda \Lambda}{R}  \right)\delta  R + \frac{ P_\Lambda}{R}  \delta (R \Lambda)  \right) \nonumber \\
	&= \int_{\mathbb R^+_0} dr \left( R^2 \delta \left( \frac{1}{2R} \left( \frac{\Lambda P_\Lambda}{R} - P_R\right) \right) + (R \Lambda) \delta \left( - \frac{P_\Lambda}{R}\right) \right) + \delta(...) \nonumber\\
	&=: \int_{\mathbb R^+_0} dr \left( E^x \delta K_x + E^\phi \delta K_\phi  \right) + \delta(...) \text{.}
\end{align}
The constraints read
\begin{align}
	H[N] &= 8\pi \int_{\mathbb R^+_0} dr \, N \left( -\frac{K_\phi^2 E^\phi}{2 (8 \pi)^2 \sqrt{E^x}} - \frac{2 K_\phi K_x \sqrt{E^x}}{(8\pi)^2} - \frac{E^\phi}{2 \sqrt{E^x}} + \frac{(E^x{}')^2}{8\sqrt{E^x} E^\phi} + \frac{\sqrt{E^x} E^x{}''}{2 E^\phi} - \frac{E^x{}' E^\phi{}' \sqrt{E^x}}{2 (E^\phi)^2}\right)				\\
	C_r[N^r] & =  \int_{\mathbb R^+_0} dr \left( E^x \mathcal L_{\vec N} K_x + E^\phi \mathcal L_{\vec N} K_\phi \right)	\text{.}
\end{align}
$K_x$ transforms with density weight $+1$, while $K_\phi$ transforms as a scalar. Consequently, $E^x$ has density weight $0$, and $E^\phi$, which coincides with $\tilde R$, $+1$. 
While this last set of variables may be considered closest to full Ashtekar-Barbero variables, the second set of variables, $\tilde P_R, \tilde R, \tilde P_\Lambda, \Lambda$, is best suited to compare to our full theory variables.

\subsection{Relation between reduced and full theory variables}

$\tilde P_R(r)$ can be reconstructed from the full theory variables as
\be
	\tilde P_R(r) = \frac{P_R(r)}{\Lambda(r)} =\frac{2}{R(r) \Lambda(r)} \int_{S^2} d^2 \theta \, { P^{AB} q_{AB}} (r,\theta) = \frac{2}{\tilde R(r)} \int_{S^2} d^2 \theta \, { \tilde P^{AB} \tilde q_{AB}} (r,\theta) 
\ee
with
\be
	\tilde R(r) = \sqrt{\frac{1}{4 \pi} \int_{S^2} d^2 \theta \, \sqrt{\det \tilde q_{AB}}} \text{.}
\ee
In order to be able to represent $\tilde P_R$ as an operator on the full theory Hilbert space, we need to rewrite it in terms of holonomies and fluxes. Using \eqref{eq:DefOfKP},

\be
	\tilde P_R(r) = - \frac{1}{\tilde R(r)} \int_{S^2} d^2 \theta \, \sqrt{\det \tilde q_{CD}} \stackrel{(\tilde P)}{K}_{AB} \tilde q^{AB}  (r,\theta) \text{.} \label{eq:PRPoissonIdentityK}
\ee
We are thus in a position employ the Poisson bracket tricks of \cite{ThiemannQSD4} which allow us to express the integrated ``would-be'' extrinsic curvature in \eqref{eq:PRPoissonIdentityK} as a Poisson bracket of the volume operator and another operator which is structurally identical to the Hamiltonian constraint in Euclidean $2+1$ gravity, but should not be confused with the proper Hamiltonian $H[N]$. We define
\be
{H}:= {F}^i_{AB}\, n^i \epsilon^{AB}
\ee 
where $F_{AB}^i = 2 \partial_{[A} A_{B]}^i + \epsilon^{ijk} A_{Aj} A_{Bk}$ is the curvature of the connection $A$. The normal $n^i$ is defined as
\bq 
n^i =\frac{\epsilon^{ijk} E^A_j E^B_k \epsilon_{AB}}{\|\epsilon^{ijk} E^A_j E^B_k \epsilon_{AB}\|}
\eq
We then find
\be
\label{eqn:po_ham}
\int_{S^2_r}\d^2\theta\,\sqrt{\det \tilde{q}_{AB}} \,\tilde{K}=\frac{\Lambda(r)}{2\beta^2}\int_{S^2_r}\d^2\theta\, \{{H},V({\cal R})\}
\ee
where $V({\cal R})$ is the 3-volume of a region ${\cal R}$ containing $S^2_r$,
\be
V({\cal R})= \int_{\cal R} dr d^2\theta \, \sqrt{\det q_{ab}} 
= \int_{\cal R} dr d^2\theta \, \Lambda^{-1} \sqrt{\det \tilde{q}_{AB}}.
\ee

\section{Quantum theory} \label{sec:QuantumTheory}

\subsection{Reduced quantisation}

In this section, we will briefly sketch the quantisation of the classically reduced sets of variables discussed in section \ref{sec:ChoicesOfReducedVariables}. Due to the agreement of the $\tilde P_\Lambda, \Lambda$ sector between the reduced and full theory, the reduced quantisation of this sector will also be valid for the full theory. We will use the quantisation rule $\{ A, B\} \rightarrow \frac{1}{i } [\hat A, \hat B]$ and work in Planck units $\hbar = c = 8 \pi G = 1$.

\subsubsection{$\tilde P_R, \tilde R, \tilde P_\Lambda, \Lambda$}

\label{sec:ReducedTildeP_RR}

We start with $\tilde P_\Lambda, \Lambda$. Due to the density weights, we integrate $\Lambda$ over an interval $e^r_i := [r_{i,1}, r_{i,2}] \subset \mathbb R^+_0$ and define 
\be
	\Lambda(e^r) := \int_{r_{1}}^{r_{2}} dr \, \Lambda(r)  \text{.}
\ee
The regularised Poisson brackets read
\be
	\left\{\Lambda(e^r), \tilde P_\Lambda (r') \right\} =  s(e^r, r') := \sign (r_{2} - r_{1}) ~ \times ~ \begin{cases} 
												1 &\mbox{if } r' \in e^r \backslash \partial e^r \\ 
												1/2 & \mbox{if } r' \in \partial e^r \\  
												0 & \mbox{otherwise.}					
											\end{cases} 
\ee
The factor $1/2$ if $r'$ is on the boundary of the interval $e^r$ originates from regularising the delta distribution, and ensures consistency when splitting the interval. Following the strategy for quantising point particles as given in \cite{ThiemannKinematicalHilbertSpaces} and applied in \cite{BojowaldSymmetryReductionFor} to the case of spherical symmetry, we exponentiate $\tilde P_\Lambda$ and define the point holonomies
\be
	h_r^\rho (\tilde P_\Lambda) = \exp \left(- i \, \rho \, \tilde P_\Lambda(r) \right)
\ee
for $\rho \in \mathbb R$. It follows that
\be
	\left\{\Lambda(e^r), h_{r'}^\rho (\tilde P_\Lambda)\right\} = - i ~ \rho ~ h_{r'}^\rho (\tilde P_\Lambda) ~ s(e^r,r')
\ee
A cylindrical function $\Psi(h^{\rho_1}_{r_1}(\tilde P_\Lambda), \ldots, h^{\rho_N}_{r_N}(\tilde P_\Lambda))$ is now defined as a function depending on a finite number of point holonomies. The space is thus spanned by `radial spin network functions' $T_{R_N, \rho_N}$ labeled by a `graph' $R_N:=\{r_1,\ldots,r_N\}$, which is coloured by $\rho_N:=\{\rho_1,\ldots, \rho_N\}$.
Given two radial spin network functions $T_1$ and $T_2$, we can reexpress both (by adding trivial, i.e. $\rho=0$, dependency) on a common refinement $\{r_1, \ldots, r_M\}$ which contains both point sets $\{r_1, \ldots, r_{N_1}\}$ and $\{r_1, \ldots, r_{N_2}\}$ on which $T_1$ and $T_2$ have support.
We then introduce the scalar product
\be
	\braket{T_1}{T_2} := \prod_{i=1}^M \delta_{\rho_i^1, \rho_i^2} \text{,} \label{eq:ScalarProductBohr}
\ee
familiar from loop quantum cosmology \cite{AshtekarMathematicalStructureOf}. The resulting Hilbert space per radial point is that of the square integrable functions (with respect to \eqref{eq:ScalarProductBohr}, i.e. $\braket{\Psi}{\Psi}<\infty$) on the Bohr compactification of the real line. The action of the operators $\hat \Lambda(e^r)$ and $\hat{h}_{r}^\rho (\tilde P_\Lambda) $ is given by 
\begin{align}
	\hat{h}_{r}^\rho (\tilde P_\Lambda) ~ \Psi(h^{\rho_1}_{r_1}(\tilde P_\Lambda), \ldots, h^{\rho_N}_{r_N}(\tilde P_\Lambda)) &~=~ h_{r}^\rho (\tilde P_\Lambda) ~ \Psi(h^{\rho_1}_{r_1}(\tilde P_\Lambda), \ldots, h^{\rho_N}_{r_N}(\tilde P_\Lambda))\\
	\hat \Lambda(e^r)  ~ \Psi(h^{\rho_1}_{r_1}(\tilde P_\Lambda), \ldots, h^{\rho_N}_{r_N}(\tilde P_\Lambda)) &~=~   \sum_{i=1}^N s(e^{r_i}, r) \, \rho \, h^{\rho_i}_{r_i} \frac{\partial}{\partial h^{\rho_i}_{r_i}} ~ \Psi(h^{\rho_1}_{r_1}(\tilde P_\Lambda), \ldots, h^{\rho_N}_{r_N}(\tilde P_\Lambda))  \text{.}
	\label{eq:lambda_operator}
\end{align}

The construction of the Hilbert space and representation for $\tilde P_R$, $\tilde R$ works in principle analogously, where $\tilde R$ takes the place of $\Lambda$ and $\tilde P_R$ the place of $\tilde P_\Lambda$. 
However, we will consider two possible changes here which are of conceptual importance when trying to relate the theories reduced at the classical and quantum level. First, since the $\tilde R, \tilde P_R$ sector in the full theory will be quantised using the gauge group\footnote{While SU$(2)$ is a gauge group in the full theory, the groups U$(1)$ and $\mathbb R_{\text{Bohr}}$ are not in the reduced context, as there is no associated Gau{\ss} law generating a gauge transformation.} SU$(2)$, we can also decide to choose U$(1)$ instead of $\mathbb R_{\text{Bohr}}$ to mimic the countable set of representation labels of SU$(2)$. The only difference to the above treatment is then that the representation labels $\rho$ are restricted to be integers. 

Another possible change is to introduce a free parameter in the spirit of the Barbero-Immirzi parameter, corresponding to $\beta$ in section \ref{sec:FullTheoryCanonicalVariables}. In order not to confuse the two, we will call the free parameter in the classically reduced theory $\gamma$. Instead of the rescaling procedure from section \ref{sec:FullTheoryCanonicalVariables}, we can simply introduce $\gamma$ as
\be
	h_{r,\gamma}^\rho (\tilde P_R) = \exp \left(- i \, \rho \gamma \, \tilde P_R(r) \right) \text{,}
\ee
so that 
\be
	\left\{\tilde R(e^r), h_{r', \gamma}^\rho (\tilde P_R)\right\} = - i ~ \rho ~ \gamma ~ h_{r', \gamma}^\rho (\tilde P_R) ~ s(e^r,r')
\ee
and an analogous quantisation follows. 

\subsubsection{$ P_R,  R,  P_\Lambda, \Lambda$}
\label{sec:ReducedP_RR}

Representation and Hilbert space for $\Lambda, P_\Lambda$ can be constructed as in section \ref{sec:ReducedTildeP_RR}. 
We again would like to polymerise $P_R$ and consider cylindrical functions depending radial holonomies of $P_R$, defined as
\be
	h^\rho_{e^r}(P_R) := \exp \left( - i \rho\int_{r_{1}}^{r_{2}} dr' \, P_R(r') \right)  \text{.}
\ee
A cylindrical function is given by $\Psi(h^{\rho_1}_{e^{r}_1}( P_R), \ldots, h^{\rho_N}_{e^{r}_N}( P_R))$ and the basis is again given by spin network functions of the same type discussed in the previous section. Given two spin-network functions we can again find a common refinement and introduce the scalar product
\be
	\braket{T_1}{T_2} := \prod_{i=1}^M \delta_{\rho_i^1, \rho_i^2} \text{.} \label{eq:ScalarProductBohr2}
\ee
The action of the operators $\hat R(r)$ and $\hat{h}_{e^{r}}^\rho ( P_R) $ is given by 
\begin{align}
	\hat{h}_{e^r}^\rho ( P_R) ~ \Psi(h^{\rho_1}_{e^{r}_1}( P_R), \ldots, h^{\rho_N}_{e^{r}_N}( P_R)) &~=~ h_{e^r}^\rho ( P_R) ~ \Psi(h^{\rho_1}_{e^{r}_1}( P_R), \ldots, h^{\rho_N}_{e^{r}_N}( P_R))\\
	\hat R(r) ~ \Psi(h^{\rho_1}_{e^{r}_1}( P_R), \ldots, h^{\rho_N}_{e^{r}_N}( P_R)) &~=~   \sum_{i=1}^N s(e^{r}_i, r) \, \rho\, h^{\rho_i}_{e^{r}_i} \frac{\partial}{\partial h^{\rho_i}_{e^{r}_i}} ~ \Psi(h^{\rho_1}_{e^{r}_1}( P_R), \ldots, h^{\rho_N}_{e^{r}_N}( P_R)) \text{.}
\end{align}

\subsubsection{$E^x, K_x, E^\phi, K_\phi$}

\label{sec:Reducedxphi}

The construction of Hilbert space and representation for $E^\phi = \tilde R$ and $K_\phi$ works analogously as in section \ref{sec:ReducedTildeP_RR} with $\tilde P_\Lambda \rightarrow K_\phi$. Since we want $E^x = R^2$ to have diagonal action on cylindrical functions, we construct the representation for $E^x, K_x$ as in section \ref{sec:ReducedP_RR} with the substitutions $P_R \rightarrow E^x$ and $R \rightarrow K_x$.

\subsubsection{Spectral properties of the reduced quantisation} \label{sec:ReducedSpectralProperties}

The spectral properties of the classically reduced quantum operators strongly depend on our choice of group. While the above discussion has been made with the choice $\RB$, we could also choose U$(1)$, which would result in the representation labels $\rho$ being integers. Consequently, the spectrum of, e.g., $\tilde R$, would not be $\mathbb R$, but only a multiple of $\mathbb Z$. This choice of group is directly tied to a classical choice of observables in the following sense: by restricting e.g. $\rho$ in $h_r^\rho (\tilde P_\Lambda)$ to be an integer, we effectively put a cut-off on $\tilde P_\Lambda$ at the order of the Planck scale. This necessarily also happens in the full theory, where one chooses SU$(2)$ as a gauge group.

With an eye on the full theory area operator to be defined later, one can ask which choice corresponds more closely to this case. In case that the operator corresponding to $\tilde R$ is the area operator and acts on a quantum reduced state which is build from a single holonomy, it seems more natural to choose U$(1)$, which mimics the discrete eigenvalues of the area operator and equal spacing for large quantum numbers in this case. On the other hand, allowing an arbitrary refinement of the quantum reduced states, $\RB$ seems more appropriate since the spectrum of the area operator becomes dense in $\mathbb R$ for large eigenvalues \cite{ThiemannModernCanonicalQuantum}. We conclude that both choices capture {\it some} features of the full theory description, but neither captures all of them. From a dynamic point of view, $\RB$ could be preferred because it allows to incorporate the $\bar \mu$-scheme of loop quantum cosmology \cite{AshtekarQuantumNatureOf}, which might also be relevant in the case of spherical symmetry.

\subsection{Full theory quantisation}

\label{sec:FullQuantisation}

\subsubsection{Quantum kinematics}

The full phase space of general relativity in the gauge fixing $q_{rA} = 0$ and $\partial_A q_{rr}=0$, as developed in section \ref{sec:Classical}, has been coordinatised by $\tilde P_\Lambda(r), \Lambda(r), E^{A}_i(r,\theta), A_A^i(r,\theta)$. The quantisation of the variables $\tilde P_\Lambda(r), \Lambda(r)$ can be copied verbatim from section \ref{sec:ReducedTildeP_RR}. For the ``angular'' variables $E^{A}_i(r,\theta), A_A^i(r,\theta)$, we follow \cite{BLSI, BLSIII} and construct holonomies and fluxes as
\begin{align}
	h^j_e(A) &:= \mathcal P \exp \left( \int_e A_{Ai}(r,\theta) \tau^i dx^A \right)   \\
	 E_m(S) &:= \int_S  E^{A}_i(r,\theta) m^i(r,\theta) \epsilon_{AB} \, dr \wedge dx^B \text{,}
\end{align}
where $\tau^i$ are the generators of the Lie algebra su$(2)$ in the representation $j$, and $m^i$ is a smearing function. Due to the tensor structure of the canonical variables, the holonomies of $A_A^i$ can be restricted to edges of constant radius $r$, whereas the surfaces over which $E^{A}_i$ is smeared should be foliated by radial geodesics. For smearing surfaces with infinitesimal radial extend, the resulting holonomy-flux algebra thus behaves effectively like a family of $2+1$-dimensional holonomy-flux algebras along the radial coordinate. 

Quantisation of the holonomy-flux algebra can be achieved, exactly as for Ashtekar-Barbero variables, by specifying the positive linear Ashtekar-Lewandowski functional $\omega_{\text{AL}}$, see \cite{ThiemannModernCanonicalQuantum} for a textbook treatment, and \cite{AshtekarRepresentationsOfThe, AshtekarRepresentationTheoryOf, LewandowskiUniquenessOfDiffeomorphism, FleischhackRepresentationsOfThe} for original literature. The only departure point from these treatments is the restriction of the allowed edges and surfaces as above. 

A cylindrical function $\Psi$, next to the dependence on $\tilde P_\Lambda$ as discussed in section \ref{sec:ReducedTildeP_RR}, can also depend on a finite number of holonomies $h^{j_1}_{e_1}(A), \ldots, h^{j_N}_{e_N}(A)$.
A basis state, that is a spin network function, thus is given by $T_{R_N,\gamma}(h^{\rho}_{r},\ldots, h^j_e(A),\ldots)=T_{R_N}(h^{\rho}_{r},\dots)\otimes T_{\gamma}(h^j_e(A),\dots)$ where $T_{\gamma}$ is a usual spin network function over graphs embedded into spheres with radii $r$. Note that in principle $\gamma$ can consist of several subgraphs embedded into different spheres.  Nevertheless, any cylindrical function has support only at a finite number of radial coordinates $r$. The Gau{\ss} law \eqref{eq:GaussLaw} requires to go over to gauge invariant cylindrical functions. Since there is always a finite distance between two radial coordinates where a cylindrical function has support, the Gau{\ss} law can only be satisfied if the cylindrical function is already gauge invariant at each radial coordinate separately, which means that gauge invariant cylindrical functions can be build from standard spin networks embedded into the spheres $S^2_r$ of constant $r$. 

The scalar product between two cylindrical functions follows directly from the more general construction for Ashtekar-Barbero variables and from \eqref{eq:ScalarProductBohr}. Given two spin network functions $T_1$ and $T_2$, we reexpress them on a common refinement $(R_M,\gamma)$ of their underlying graphs $\gamma_1, \gamma_2$ and their radial intervals $R_{N_1}$ and $R_{N_2}$. The scalar product then simply reads
\begin{align}
 	\braket{T_1}{T_2} & =\braket{T^1_{R_M}}{T^2_{R_M}} \times \int_{\text{SU}(2)} d g_1 \ldots d g_{N}  \,\overline{T^1_{\gamma}\left(g_1, \ldots, g_N \right)} \, T^2_{\gamma} \left(g_1, \ldots, g_N \right)  \text{,}
\end{align}
where $1, \ldots, N$ label the edges of $\gamma$ and  $\braket{T^1_{R_M}}{T^2_{R_M}}$ is defined by \eqref{eq:ScalarProductBohr}. Thus, the scalar product reduces to the usual LQG product for cylindrical functions that do not depend on $\tilde{P}_{\Lambda}$. 

The fluxes $E_m(S)$ can be quantised exactly in the same manner as in full LQG  with the restriction of the surfaces to be foliated by radial geodesics. The resulting flux operator acts as usual like a right invariant vector field at the intersection points of the smearing surface $S$ with edges of the spin network. For a detailed record see \cite{ThiemannModernCanonicalQuantum} and \cite{BLSIII} in the case of a radial gauge fixing.  

On top of the Gau{\ss} law, we have the radial diffeomorphism constraint \eqref{eq:DiffConstrRadConnection} with constant shift vectors, which can be solved by going over to radially diffeomorphism-invariant states, following the ideas of \cite{AshtekarQuantizationOfDiffeomorphism}. Since we effectively only deal with one dimension in the case of radial diffeomorphisms with angle-independent shift vectors, the problem reduces to the one e.g. studied in \cite{BojowaldSymmetryReductionFor} in the context of spherical symmetry, and many of the subtleties discussed in \cite{AshtekarQuantizationOfDiffeomorphism} are not present. The radial diffeomorphism constraint can be explicitly solved by a simple rigging map.
Following \cite{AshtekarQuantizationOfDiffeomorphism}, we define 
\be
	\Psi(h^{j_1}_{e_1}(A), \ldots ; h^{\rho_1}_{r_1}(\tilde P_\Lambda), \ldots) ~ \mapsto ~ \eta[\Psi] :=\!\!\!\sum_{\phi \in \text{Diff}_{\text{}}(\mathbb R^+_0) / \text{Diff}_\gamma} \!\!\!\Psi^*(h^{j_1}_{\phi(e_1)}(A), \ldots; h^{\rho_1}_{\phi(r_1)}(\tilde P_\Lambda), \ldots) ,
\ee
where $\Psi^* = \braket{\Psi}{\cdot}$ is the algebraic dual of $\Psi$ and $\text{Diff}_\gamma$ are the diffeomorphisms preserving the graph $\gamma$ underlying $\Psi$. The scalar product between two diffeomorphism-invariant states is defined as
\be
	\braket{\eta(\Psi_1)}{\eta(\Psi_2)}_{\text{Diff}} := \eta(\Psi_1)[\Psi_2] \text{.}
\ee

\subsubsection{Geometric operators} \label{sec:GeometricOperators}

\paragraph{Area operator:} The area operator is constructed in the same way as for $3+1$-dimensional loop quantum gravity \cite{SmolinRecentDevelopmentsIn, AshtekarQuantumTheoryOf1}. Since our fluxes are restricted to surfaces which are foliated by radial geodesics, also the area operator is restricted to those surfaces. Up to this difference, the operator is identical to the usual one, in particular it has the same spectrum and action on spin network states.

\paragraph{Volume operator:}

\label{sec:VolumeOperator}

The volume operator \cite{AshtekarDifferentialGeometryOn} plays a central role in the definition of the dynamics of loop quantum gravity, where it enters through Poisson bracket identities \cite{ThiemannQSD1} defining phase space functions which are not easily derivable from holonomies and fluxes. We will use similar techniques in this paper, following the discussion in \cite{BLSI, BLSIII}. For self-consistency, we will again sketch the construction of the corresponding volume operator, which differs only marginally from the one defined in \cite{ThiemannQSD4, BLSI}. The main difference to the well-known volume operator in $3+1$ dimensions is that our operator, due to its construction from effectively $2+1$-dimensional holonomy-flux variables, has a non-vanishing action of gauge-invariant three-valent vertices, and even gauge invariant non-trivial two-valent vertices, that is kinks in a holonomy.

However, the volume operator here is still three dimensional, which expresses itself through the dependence on $\Lambda$. It is exactly this dependence which forbids a direct application of the point-splitting regularisation used in \cite{ThiemannQSD4}. But note that at fixed $r$ $\det \q_{AB}$ is formally identical to the volume form in $2+1$-gravity. Thus, the strategy in the following will be to first regularise the $r$-direction and then to apply point-splitting to the angular part. This strategy will also prove to be very useful when regularising $\PR$.  

We start from the classical expression 
\be
	V(R) = \int_R dr \, d^2 \theta \, \sqrt{\det q_{ab}} = \int_R dr \, d^2 \theta \, \sqrt{\Lambda^2 \det q_{AB}} = \int_R dr \, d^2 \theta \, \sqrt{ \frac{ V^k  V_k }{\Lambda^2}}
\ee
of the volume $V(R)$ of a three-dimensional region $R$ with 
\be
	 V^k  := \beta^2 \frac{ E^{A}_i  E^{B}_j \epsilon_{AB} \epsilon^{ijk} }{2} \text{.}
\ee
To regularise the radial integral, we divide the radial part of the region ${\cal R}:=[r_0,r_1]\times {\cal R}_{\Theta}$ into intervals $e^{\epsilon}_{r_i}:=[r_i-\epsilon/2,r_i+\epsilon/2]$ of coordinate length $\epsilon$. For $\epsilon$ small enough, we thus obtain the following approximation:
\begin{align} \nonumber
V({\cal R}) &=\sum_{i} \int_{e^{\epsilon}_{r_i}} dr\int_{{\cal R}_{\Theta}} d^2\theta\, \sqrt{ \left(\frac{V^k(r,\theta)}{\Lambda(r)}\right)^2} 
=\sum_{i} \epsilon \int_{{\cal R}_{\theta}} d^2\theta\, \sqrt{ \left(\frac{V^k(r_i,\theta)}{\Lambda(r_i)}\right)^2}
+{\cal O}(\epsilon^2) 
\\
\label{eq:approxW}
&= \sum_{i}\int_{{\cal R}_{\theta}} d^2\theta\, \sqrt{ \left(\frac{V^k(e^{\epsilon}_{r_i},\theta)}{\Lambda(e^{\epsilon}_{r_i})}\right)^2}
+{\cal O}(\epsilon^2) := \sum_{i}  \frac{W^{\epsilon}_{r_i}({\cal R}_{\theta})}{|\Lambda(e^{\epsilon}_{r_i})|} +{\cal O}(\epsilon^2) ~,
\end{align}
where in the second to last step we used the approximation
\begin{align}
\label{eq:V_epsilon}
\epsilon^2 V^k(r_i, \theta) 
= \beta^2 \frac{E^{A}_i(e^{\epsilon}_{r_i},\theta)  E^{B}_j(e^{\epsilon}_{r_i},\theta) \epsilon_{AB} \epsilon^{ijk} }{2}+{\cal O}(\epsilon^2)
:= V^k(e^{\epsilon}_{r_i},\theta)+{\cal O}(\epsilon^2)
\end{align}
and $\epsilon\Lambda(r_i)=\Lambda(e^{\epsilon}_{r_i})+{\cal O}(\epsilon^2)$ with $ E^{A}_i(e^{\epsilon}_{r_i},\theta):=\int_{e^{\epsilon}_{r_i}}dr\,E^{A}_i(r,\theta)$ and $\Lambda(e^{\epsilon}_{r_i}):=\int_{e^{\epsilon}_{r_i}}dr\,\Lambda(r)$. Since $\Lambda(e^{\epsilon}_{r_i})$ only acts on the $\Lambda$-component and $W^{\epsilon}_{r_i}({\cal R}_{\theta})$ only on the angular component of a cylindrical function, they can be quantised independently. Note that the operator $\widehat{\Lambda(e^{\epsilon}_{r_i})}$ defined by \eqref{eq:lambda_operator} generically contains zero in its spectrum. Therefore, its inverse is not well-defined. In order to give meaning to an operator $\widehat{|\Lambda(e^{\epsilon}_{r_i})^{-1}|}$, one can, however, apply another Poisson-bracket trick, that is replace $\Lambda^{-1}$ by $(2\{\sqrt{\Lambda},\tilde{P}_{\Lambda}\})^2$. But since in the final expression of $\PR$ all $\Lambda$ factors cancel, we will not further elaborate on this issue and instead concentrate on the quantisation of $W^{\epsilon}_{r_i}({\cal R}_{\theta})$.\footnote{If we would regularise the $\Lambda$ terms in each individual piece, we would obtain an operator where the $\Lambda$-dependence would be suppressed at least in the large quantum number limit, which we will be interested in in the end.}

Without loss of generality, taking $\epsilon$ small enough we might assume that only those spin networks of an arbitrary cylindrical function lie in the support of the operator corresponding to $W^{\epsilon}_{r}({\cal R}_{\theta})$ that are defined over graphs on the shell $S^2_{r}$. In this case, $W^{\epsilon}_{r}({\cal R}_{\theta})$ is formally equivalent to the 2+1 volume. Moreover, the Poisson bracket of $E^{A}_j(e^{\epsilon}_{r_i},\theta)$ and $A^j_{A}$ reduces effectively to 2 dimensions, that is  $\{\A^i_A(r,\theta),\E^B_j(e^{\epsilon}_r, \theta')\}= \delta^{(2)}(\theta,\theta') \delta^I_j \delta^A_B$. Consequently, $W^{\epsilon}_r$ can now be quantised along the lines of  \cite{ThiemannQSD4}. The final action of the operator $\widehat{W_r^\epsilon({\cal R}_{\theta})}$ on a spin network over $\gamma_r\subset S^2_r$ is thus given by 
\be 
\label{volumeoperator}
\widehat{W_r^\epsilon({\cal R}_{\theta})} f_{\gamma_r}= \frac{\beta^2}{4} \sum_{v\in V(\gamma_r,{\cal R}_{\theta})} \sqrt{
\left(\frac{1}{2}\sum_{e, e'\in E(\gamma_r,v)} \mathrm{sgn}(\det(\dot{e},\dot{e}'))\, \epsilon_{ijk} X^j_e X^k_{e'}\right)^2}\,f_{\gamma_r}~,
\ee
where $V(\gamma_r,{\cal R}_{\theta})$ is the set of vertices in $\gamma_r$ lying inside the region ${\cal R}_{\theta}$ (for fixed $r$), $E(\gamma_r,v)$ is the set of edges of $\gamma_r$ intersecting in $v$, and $X^j_e$ are right invariant vector fields at $v$  along $e$ defined by $X^i_e(t)= \mathrm{tr}\left[h_e(0,t)\tau^i h_e(t,1)\right]^T \frac{\partial\hphantom{h_e}}{\partial h_e(0,1)}$, where $t=0$ if $e$ is outgoing from $v$ and $t=1$ if $e$ is ingoing. Note that the right hand side of \eqref{volumeoperator} is independent of $\epsilon$ so that in the following we can suppress this label. The full volume operator is then given by:
\bq
\widehat{V({\cal R})}\Psi
=\sum_{r\in {\cal R}} \widehat{|\Lambda(e^{\epsilon}_r)^{-1}|} \,  \widehat{W_r({\cal R}_{\theta})} \Psi_r 
\eq
where by the operator corresponding to $1/\Lambda$ we mean a suitable regularisation as discussed above.

\subsubsection{Spectral properties} \label{sec:FullSpectralProperties}

A first comparison between the full and reduced theories can be made at the level of geometric operators. In section \ref{sec:ReducedSpectralProperties}, we saw that in the $\tilde R, \tilde P_R$ variables, which are closest to our full theory variables, $\hat {\tilde R}(e^r)$ acts on quantum states as
\be
\label{425}
	\hat {\tilde R}(e^r) \ket{\exp \left(- i \, \rho \gamma \, \tilde P_R(r) \right) } = \gamma \rho \ket{\exp \left(- i \, \rho \gamma \, \tilde P_R(r) \right) } \text{,}
\ee
i.e. via multiplication of $\gamma \rho$. To this, we can first compare the action of the area operator in the full theory. For this comparison, we take as our full theory quantum state a Wilson loop at radial coordinate $r$ with representation label $j$. The area operator we consider is defined for an area intersecting this loop twice. For example, we could take a great circle on $S^2_r$ times an interval in radial direction, which corresponds, up to a numerical coefficient, to ${\tilde R}(e^r)$, and intersect this area with a holonomy defined along another great circle. On such a state, here simply denoted by $\ket{ j}$, the area operator, here denoted by $\hat A_r$, acts as
\be
	\hat A_r \ket{ j} = 2 \beta \sqrt{j(j+1)}  \ket{ j} \text{.}
\ee
We could thus decide to specify our mapping between reduced and full theory states by setting the ratio of $\gamma$ and $\beta$ such that we can map $\rho = j$ at least approximately for large quantum numbers. Unfortunately, there are several problems with this. 
\begin{itemize}
	\item The above construction using the full theory area operator strongly depends on the details of how the full theory quantum state is specified, i.e. the intersection properties of the holonomy with the area specifying the area operator. It does not seem that the details can be made invariant under spatial diffeomorphisms preserving the $S^2_r$, which is however our reduction procedure to be carried out. We could circumvent this by choosing a different reduction procedure by averaging only over rigid rotations instead of spatial diffeomorphisms, as discussed in \cite{BLSI}, however the regularisation of the full theory quantum operator for $ \tilde P_R$ is then not any more well motivated. 
	\item The above identification at large quantum numbers does not work for the volume operator acting on the kink state that we will discuss later in the context of the quantum reduction. While one would naively expect that the action of the volume operator is proportional to $j^2$ at large quantum numbers, the terms proportional to $j^2$ cancel and the operator is proportional to $j$ for large $j$, which gives $\tilde R \sim \sqrt{j}$ asymptotically. This problem can however be made to disappear to some extend by using a regularisation where the volumes act mostly on non-gauge-invariant kink states, as discussed later. It thus has to be attributed to the simplicity (degeneracy) of the kink state that we are using. 	
	\item The scaling problems of the volume operator should disappear as soon as one chooses at least three-valent vertices. However, besides of the action of the volume becoming more complicated then (since due to non-gauge-invariance effects, the vertex is effectively four-dimensional in several steps of the computation), it is unclear how to precisely map a collection of $j$ labelling a full theory quantum state to the single quantum number $\rho$ labelling the reduced state. 
	\item Last but not least, the spectral properties between the reduced and full theory don't seem to match, as discussed in section \ref{sec:ReducedSpectralProperties}.
\end{itemize}
In conclusion, there is so far no satisfactory way to map the reduced theory quantum states to full theory quantum states such that the spectral properties of the operators we are considering are roughly the same and that all steps in the computations are well motivated. However, it transpires that at large quantum numbers, one should identify $\rho = j$ for a suitable choice of the ratio of $\beta$ and $\gamma$, the free Barbero-Immirzi-type parameters of the full and reduced theories. 

\subsubsection{Imposition of reduction constraints}

We are now in a position to discuss the reduction condition $\left. C_A[N^A] \right|_{q_{rB}=P^{rB}=0} = 0$, as defined in equation \eqref{eq:DiffConstrAngConnection}. Again following \cite{AshtekarQuantizationOfDiffeomorphism}, we implement it as invariance under finite diffeomorphisms, which result from exponentiating the action of \eqref{eq:DiffConstrAngConnection}. The solution is again given by diffeomorphism equivalence classes of (angular) spin networks, however now also the issue of graph symmetries \cite{AshtekarQuantizationOfDiffeomorphism} comes into play. Since a cylindrical function $\Psi$ has support only at a finite number of radial coordinates, we can always find vector field $N^A$ such that the action of \eqref{eq:DiffConstrAngConnection} preserves $A_{a}^{i}$ at all but one of the radial coordinates on which $\Psi$ has support. It follows that the problem reduces to the $2+1$-dimensional one, which has been sketched in \cite{ThiemannQSD4}, just that here we consider the special case of a spherical topology.
We emphasise that implementing \eqref{eq:DiffConstrAngConnection} $=0$ as the action of finite angular spatial diffeomorphisms has to be regarded as the reduction step, whereas the steps before still apply to full general relativity.

\subsubsection{Regularisation of $\PR$}

In order to define an operator corresponding to $\PR$ in the full quantum theory, we will first have to absorb all inverse powers of $\sqrt{\det{\q_{AB}}}$ inside Poisson brackets of the volume. We already expressed $\PR$ in terms of a bracket with the `Hamiltonian' ${H}$. It remains to find an adequate expression for $n^i$, 
for which we follow \cite{ThiemannQSD4}. The normal $n^i$ can be rewritten as
\bq
n^i= \frac{2}{\beta^2}\int_{\cal R} dr d^2 \theta \int_{\cal R} dr' d^2 \theta' \, \epsilon^{ijk} \epsilon^{AB} \{A^j_A,(\det\q_{A'B'}(r,\theta))^{1/4}\}\, \{A^k_B,(\det\q_{A'B'}(r',\theta'))^{1/4}\} \text{.}
\eq
The strategy is now to introduce a point splitting regularisation similar to \cite{ThiemannQSD4} such that the above expression is replaced by 
\bq 
n^i= \frac{2\Lambda}{\epsilon \beta^2} \,\epsilon^{ijk} \epsilon^{AB} \{A^j_A,\sqrt{W({\cal R}_{\epsilon})}\}\, \{A^k_B,\sqrt{W({\cal R}_{\epsilon})}\} +{\cal O}(\epsilon^4) \text{,}
\eq
where ${\cal R}_{\epsilon}$ is a region of coordinate volume $\epsilon^3$ and $W({\cal R})=\int_{\cal R} dr d^2 \theta \,\sqrt{\det\q_{AB}}$. The additional factor $\epsilon^{-1}$ appears here since $\sqrt{\det\q_{AB}}$ does only transform as a density in angular directions, but not in the radial one. This, at first sight, seems to be worrisome because it seems to lead to an operator that is not well-defined in the limit as $\epsilon$ goes to zero. However, the factor $\epsilon^{-1}$ can be absorbed when first regularising the radial direction of the full expression $\PR$ similar to the regularisation of the volume as previously discussed in section \ref{sec:GeometricOperators}. \footnote{Another option would be to replace $W$ by $V$ and multiply by $\Lambda$, but then one would need to quantise the inverse of $\Lambda$ in $V$.}.

Consider again a region of type ${\cal R}_r^{\epsilon}= e_r^{\epsilon}\times S^2_r$ with $e_r^{\epsilon}=[r-\epsilon/2, r+\epsilon/2]$ and define 
\begin{align}
n_i^{\epsilon}(r,\theta):= \frac{\epsilon_{ijk}\,\epsilon^{AB}}{2\beta^2}  \int_{{\cal A}^{\epsilon}_r} \d^2\theta_1\d^2\theta_2
\frac{ \{\A^j_A(r,\theta),  \sqrt{ (V^m(e_r^{\epsilon},\theta_1))^2}\,\}}{(( V^m(e_r^{\epsilon},\theta_1)^2) )^{1/4}}
\frac{ \{\A^k_B(r,\theta),  {\sqrt{ (V^m(e_r^{\epsilon},\theta_2))^2 }}\,\}}{( (V^m(e_r^{\epsilon},\theta_2))^2)^{1/4}}~,
\end{align}
where ${\cal A}^{\epsilon}_r$ is a region of size $\epsilon^2$ containing the point $(r,\theta)$ and $V^m(e_r^{\epsilon},\theta)$ was defined in \eqref{eq:V_epsilon}. It follows from \eqref{eq:approxW} that 
\begin{align}
\PR(r) 
=& \nonumber
 \frac{-1}{2\beta^2 \R(r)} \int_{S^2_r} \d^2\theta \,\{\epsilon^{AB} \,{F}^i_{AB} \,n_i , W({\cal R}_r^{\epsilon})\}\\
=&
\frac{-1}{2\beta^2\,\epsilon\, \R(r)} \int_{S^2_r}\d^2\theta \left\{\epsilon^{AB} {F}^i_{AB}(r,\theta) \, n_i^{\epsilon}(r,\theta) \,,W^{\epsilon}_r(S^2_r)  \right\} \,+{\cal O}(\epsilon^2)~.
\end{align}
On the other hand, we find 
\bq
\epsilon \,\R(r)=\frac{1}{\sqrt{4\pi}} \int_{e_r^{\epsilon}} dr \left[ \int_{S^2_r} d^2\theta \,\sqrt{\det\q_{AB}}\right]^{1/2} +{\cal O}(\epsilon^2)=  \sqrt{\frac{1}{4\pi}\,W_r^{\epsilon}(S^2_r)} +{\cal O}(\epsilon^2)~,
\eq
from which it follows that the radial regularised momentum,
\be
\PR(e^{\epsilon}_r):= - \frac{\sqrt{4 \pi}}{\beta^2}\int_{S^2_r} \d^2\theta \left\{\epsilon^{AB} {F}^i_{AB}(r,\theta) \, n_i^{\epsilon}(r,\theta) \,,\sqrt{W^{\epsilon}_r(S^2_r)}  \right\}~,
\ee
converges to $\PR(r)$ in the limit $\epsilon\to 0$. By absorbing $\epsilon \,\R(r)$ in the Poisson bracket, we automatically get rid of the problem that the operator associated to $\R$ generically contains zero in its spectrum, so that its inverse is not well-defined. Furthermore, we can now immediately replace the `volume' $W^{\epsilon}_r$ by its quantum analog \eqref{volumeoperator} and apply the point-splitting procedure \cite{ThiemannQSD4}  to the remaining part.

Let $\chi_{\delta}(\theta,\theta')$ denote the characteristic function of a square\footnote{That is, $ \chi_{\delta}(\theta,\theta')$ equals one if $\delta/2-|\theta^1-{\theta'}^1|>0$ and $\delta/2-|\theta^2-{\theta'}^2|>0$ and vanishes otherwise.} with area $\delta^2$ and define
\bq
W^{\epsilon}_r(\delta,\theta)= \int\d^2\theta' \,\chi_{\delta}(\theta,\theta')\,\sqrt{(V^k(e^{\epsilon}_r,\theta'))^2}~.
\eq
Then, following \cite{ThiemannQSD4}, one finds:
\begin{gather} \nonumber
\lim_{\delta\to0} \,\int_{S^2_r}\d^2\theta\,\epsilon^{AB} {F}^i_{AB}\int_{S^2_r}\d^2\theta'
\, \chi_{\delta}(\theta,\theta')\, \epsilon^{CD}\,\epsilon_{ijk} \,\{\A^j_C(r,\theta'), \sqrt{W^{\epsilon}_r(\delta,\theta')}\}\,
 \{\A^k_D(r,\theta'), \sqrt{W^{\epsilon}_r(\delta,\theta')}\}
 \\
 =\frac{\beta^2}{2 }\int_{S^2_r} \epsilon^{AB} {F}^i_{AB} \, n_i^{\epsilon}~.
\end{gather}   
The curvature and connections can be regularised by introducing a triangulation ${\cal T}$ of the sphere $S^2_r$. In each triangle $\Delta$ of ${\cal T}$, we select a source vertex $v(\Delta)$ and label the two edges adjacent to $v(\Delta)$ by $e_1(\Delta)$ and $e_2(\Delta)$, where the numbering is induced by the orientation. That is, if $\dot{e}_1$ and $\dot{e}_2$ are both outgoing, then they satisfy\footnote{For a more precise, diffeomorphism invariant definition of the orientation see \cite{ThiemannQSD4}.} $\det(\dot{e}_1,\dot{e}_2)>0$. As usual, the connections are regularised by considering holonomies along $e_1,e_2$ and the curvature is approximated by holonomies along the loops $\alpha(\Delta)=e_1^{\epsilon_1}(\Delta)\circ a_{12}(\Delta)\circ e^{-\epsilon_2}_2(\Delta)$ where $a_{12}(\Delta)$ is the arc connecting the vertices $e_1(1)\neq v(\Delta)$ and $e_2(1)\neq v(\Delta)$, and where $\epsilon_i$ equals $1$ if $e_i$ is outgoing and $- 1$ otherwise. Thus in an arbitrary irreducible representation $k\neq0$ we have
\bq
	h^{\epsilon_I}_{e_I(\Delta)}(A)= 1+\eta\, \dot{e}_I^A\, A^i_A(v(\Delta)) \, \tau_i^{(k)}+{\cal O}(\eta^2)
\eq
and
\bq
	h_{\alpha(\Delta)}(A)=1+\frac{\eta^2}{2} \,\dot{e}_1^A \dot{e}_2^B\, F^i_{AB}(v(\Delta)) \, \tau_i^{(k)} +{\cal O}(\eta^3)~.
\eq
Here, $\eta$ is the approximate length of $e_I$ ($I\in\{1,2\}$), $\dot{e}_I^A$ is the unit tangent of the edge $s_I$ at $v(\Delta)$ and $\tau^{(k)}_i$ are the matrices corresponding to the generators of $\mathfrak{su}(2)$, e.g. $\tau^{(1/2)}_i=\frac{1}{ 2i} \sigma_i$ where $\sigma_i$ are Pauli matrices, in the irrep. $k$. On the other hand:
\bq
\tr[ \tau_i^{(k)}  \tau_j^{(k)}  \tau_k^{(k)}]=-\frac{k (k+1) (2k+1)}{3!} \epsilon_{ijk} := - C(k) \epsilon_{ijk}
\eq
and $2 \dot{e}_1^A\dot{e}_2^B F^i_{AB}= \epsilon_{AB}  \dot{e}_1^A\dot{e}_2^B \epsilon^{CD} F^i_{CD}$. Thus, decomposing $\int_{S^2_r}$ into $\sum_{\Delta} \int_{\Delta}$ and approximating $\int_{\Delta} \cdots$ through
$\dot{e}_1^A\dot{e}_2^B \epsilon_{AB} \eta^2/2 \cdots$, one finally arrives at the regularised `Hamiltonian'
\begin{align}
\label{regHAm}
\begin{split}
{H}^{\epsilon,\delta}_{\cal T} &:= -\frac{1}{\beta^2 C(k)} \sum_{\Delta,\Delta'\in{\cal T}} \epsilon^{IJ}\epsilon^{KL}\, \chi_{\delta}(v(\Delta),v(\Delta'))\\
&\times \tr\left[h_{\alpha_{IJ}(\Delta)} h^{\epsilon_K}_{e_K(\Delta')}\left\{h^{-\epsilon_K}_{e_K(\Delta')}, \sqrt{W_r^{\epsilon}(\delta, v(\Delta')})\right\} 
 h^{\epsilon_L}_{e_L(\Delta')} \left\{h^{-\epsilon_L}_{e_L(\Delta')}, \sqrt{W_r^{\epsilon}(\delta, v(\Delta'))}\right\}\right]~,
\end{split}
\end{align}
which can be easily quantised by replacing the classical functionals by operators and the Poisson brackets by $\frac{1}{i}[\cdot,\cdot]$. As usual, we will have to adapt the triangulation to a given graph such that the operator has certain properties, amongst others it should only act on the nodes of a given spin network, and such that the regulators can at least partially be removed.

\subsubsection{Design of the triangulation}

Note that \eqref{regHAm} tends to ${H}$ if we first remove the triangulation, i.e. send $\eta$ to zero, then take the limit $\delta\to 0$ and finally take the limit $\epsilon\to 0$. In the quantum theory, these limits can be interchanged (see \cite{ThiemannQSD4}). Since only $W_r^{\epsilon}$ depends on $\epsilon$, but the quantum operator has the same spectrum for $\epsilon$ small enough to single out a shell $S^2_r$, which is a diff-covariant statement, we can easily remove this regulator independently of the others once we impose the radial diffeomorphism constraint\footnote{To be precise, the limit can be taken in the space of radial-diff-invariant algebraic distributions ${\cal D}_{r-\text{diff}}^{\ast}$. If $\Psi\in{\cal D}_{r-\text{diff}}^{\ast}$ and $f\in {\cal H}_{kin}$ then the limit $\lim_{\epsilon\to0} \Psi( \widehat{{H}^{\epsilon,\delta}_{\cal T}}f)$ converges uniformly to $\Psi( \widehat{{H}^{\epsilon_0,\delta}_{\cal T}} f):=\Psi( \widehat{{H}^{\delta}_{\cal T}} f)$ for some arbitrary but fixed parameter $\epsilon_0$.}. Thus, we could in principle just use the same strategy as in \cite{ThiemannQSD4} in order to get rid of the remaining regulators. However, here, we wish to construct a non-graph-changing operator so that the resulting theory becomes as similar as possible to the theory one obtains from quantising the classically reduced system (see section \ref{sec:FullSpectralProperties}). The motivation behind our regularisation thus has to be seen purely in this light. When going beyond the kink state, these prescriptions might have to be adapted.

In order to obtain a non-graph-changing operator, we have to consider a generalised triangulation of $S^2_r$ consisting of convex 2-dimensional polyhedra\footnote{In a slight abuse of notation these will be further labelled by $\Delta$ in order to keep the notation simple.} instead of triangles. As before, we will choose one base point $v(\Delta)$ in every polyhedron around which $\chi_{\delta}$ in \eqref{regHAm} will be concentrated. Since the limits $\eta\to 0$ and $\delta\to 0$ are interchangeable in the quantum theory (see \cite{ThiemannQSD4} for details), it is now possible to choose $\delta$ small enough such that the only vertex in the support of $\chi_{\delta}(v(\Delta),\cdot)$ is $v(\Delta)$ itself. In doing so, one also guarantees that $\widehat{W_r(\delta, v(\Delta'))}$ is only acting on $v(\Delta)$. Thus $[h^{-1}_{s_L(\Delta')}, \sqrt{W_r(\delta, v(\Delta'))}]$ is only non-zero if $v(\Delta)$ lies on the graph $\gamma$ underlying the spin network on which $\hat{{H}}$ acts. Since in contrast to the usual volume operator in 3+1 gravity, $\hat{W_r}h_s$ does not vanish if $s$ intersects $\gamma$ not in a vertex but inside an edge, we are forced to adapt the triangulation ${\cal T}$ of $S^2_r$ in such a way that if a base point $v(\Delta)$ of a polyhedron in the triangulation lies in $\gamma$, then it is a vertex of $\gamma$. In order to obtain a non-graph-changing operator, we additionally require that if $v(\Delta)$ is a vertex of $\gamma$, then $\partial\Delta$ lies entirely in $\gamma$ and is a convex, minimal loop of $\gamma$. Here, a loop of $\gamma$ is called minimal if it does not contain any sub-loops. Note that in 2-dimensions a minimal loop at $v(\Delta)$ is uniquely determined by two neighbouring edges $e,e'$ for which $e\cap e'=v(\Delta)$ and $e,e'\subset\partial\Delta$. The main reason for considering only a minimal convex loop is that we aim at a formulation in which the triangulation can be refined by shrinking the loop while adding more polyhedra, so that the classically regularised expression of $\PR$ tends to the continuum version for small loops. But this is in general not possible if one admits non-convex cells. For example, a loop with a kink divides the sphere into a generalised convex and a non-convex polyhedron that cannot be shrunken simultaneously. Let us stress here that convexity in 2-dimension is invariant under orientation preserving diffeomorphisms, i.e. diffeomorphisms connected to the identity, which are considered here. This can be easily seen by approximating a convex polyhedron by an arbitrarily fine piecewise linear (p.l.) cell $\tilde{\Delta}$. For a p.l. cell there exist a labelling of the edges $e_i(\tilde{\Delta}),\,i=1,\cdots,n$ in $\partial\tilde{\Delta}$, unique up to cyclic permutations, s.t. $e_i(\tilde{\Delta})\cap e_{i+1}(\tilde{\Delta})=v_i$, where $v_i$ are the vertices of the cell, and all $\det(\dot{e}_i(\tilde{\Delta}),\dot{e}_{i+1}(\tilde{\Delta}))$ are positive. Thus in order to transform a convex cell into a non-convex cell, at least one of the determinants need to change sign, which requires an orientation changing diffeomorphism.

So far, the above description is completely analogous to the design of the triangulation in \cite{ThiemannQSD4}, except for the slight modification that $\partial \Delta$ lies entirely in $\gamma$ and only pairs of edges spanning minimal convex loops are considered. It is, however, the first modification that forbids to consider a triangulation in which every vertex of $\gamma$ is a base point as it is done in \cite{ThiemannQSD4}. If one would do so, one would need to consider several base points per $\Delta$, which would cause problems when point-splitting because in this case the volume can not be localised just in $v(\Delta)$. A solution is to construct a family of triangulations, one for each proper vertex $v\in\gamma$, i.e. for every vertex with at least two non-collinear edges labelled by a non-trivial representation. Moreover, it is necessary to consider also a separate triangulation for each pair of edges incident at $v$ that span a minimal convex loop. For the same reason, a separate triangulation is constructed for every pair of neighbouring edges in \cite{ThiemannQSD4}.

In the following, let us explicitly describe how to construct a generalised triangulation for the vertex $v$ and the pair of neighbouring edges $(e,e')$ that span a minimal convex loop $\partial\Delta$ with base point $v(\Delta)=v$. The first step in completing the triangulation is to add $n-1$ (convex) polyhedra, where $n$ should be small, such that we obtain a triangulation of a neighbourhood $U_v(\Delta)$ of $v$ by $n$ polyhedra. Thereby, the additional polyhedra should be, at best, determined uniquely by $\Delta$, so that they transform accordingly under diffeomorphisms changing $\Delta$, and divide $U_v(\Delta)$ in $n$ regions of approximately the same coordinate size such that $\int_{U_v(\Delta)}\dots$ can be approximated by $n \int_{\Delta}\dots$. Furthermore, we are especially interested in the case where \eqref{regHAm} gives a good approximation of the continuum expression. Thus, we should assume that the coordinate size of $\Delta$ approximately equals the coordinate size of the triangle spanned by $e$ and $e'$. Note that this assumption does not cause any problems since the final operator does not depend on the size of $\Delta$ and once spherical diffeomorphisms are implemented, i.e. spherical reduction is implemented, we can always find representatives of this kind in each equivalence class. In order to obtain a triangulation, we can thus just add 3 additional triangles by the mirror construction described in \cite{ThiemannQSD4}. Yet, there is a second option to construct a triangulation for (different) $U_v(\Delta)$: Given $\Delta$, we can shoot out a geodesic $\bar s$ from $v$ with respect to the standard metric on $S^2$ such that we obtain 2 additional triangles by connecting the endpoint of $\bar{s}$ with the second vertex (not $v$)  of $e$ and $e'$ respectively. Note that $\bar s$ is fixed completely by the requirement that the coordinate size of the new triangles equals approximately the size of $\Delta$. This construction also leads to a diffeomorphism covariant operator since a rescaling of $\Delta$ also changes $\bar s$, but does not effect the standard metric on $S^2$ in this prescription. Since at this point it is not clear which of the two ways to construct $U_v(\Delta)$ should be preferred, we will keep this as a quantisation ambiguity.

To proceed with the regularisation of $\PR$, we construct a neighbourhood $U_{v,i}:=U_v(\Delta_i)$ for every convex minimal loop $\partial\Delta_i$ at $v$ and denote the union of all these neighbourhoods by $S(v)$, i.e.
\bq
S(v)=\bigcup_i^{E(v)} U_{v,i}
\eq
where $E(v)$ is the number of minimal convex loops at $v$. The complement of $S(v)$, that is $\bar{S}(v)=S^2_r - S(v)$, and $S(v)- U_{v,i}$ can now be triangulated arbitrarily under the condition that no base point of polyhedra in $\bar{S}(v)$ and $S(v)- U_{v,i}$ lies in $\gamma$, which yields a triangulation of $S^2_r$ for each $\Delta_i$. By averaging over the full family of triangulations we can decompose the integral $\int_{S^2_r}$ as follows:
\bq
\int_{S^2_r}= \int_{\bar{S}(v)} +\left[\frac{1}{E(v)} \sum_{i=1}^{E(v)} \int_{S(v)- U_{v,i}}\right]
 +\left[\frac{n}{E(v)} \sum_{i=1}^{E(v)} \int_{\Delta_i}+{\cal O}(\mathrm{Ar}(\Delta_i)^{3/2})\right]
\eq
where $n$ equals $4$ for the mirror construction \cite{ThiemannQSD4} and $3$ for the second way of constructing $U_v(\Delta)$. Note that we do not sum over the vertices in contrast to \cite{ThiemannQSD4}, because the triangulation was build explicitly just for $v$. In order not to single out a vertex in $\gamma$ one therefore has to average over all families of triangulations for all proper vertices, i.e. 
\bq
\int_{S^2_r}= \frac{1}{\|\gamma^{(0)}_{prop}\|} \sum_{v\in\gamma^{(0)}_{prop}} \left[ \int_{\bar{S}(v)}+\cdots\right]
\eq
where $\gamma^{(0)}_{prop}$ is the set of proper vertices and $\|\gamma^{(0)}_{prop}\|$ its cardinality. We here have to divide by $\|\gamma^{(0)}_{prop}\|$ since we consider for each vertex a different family of triangulations and then sum over all triangulations. This at the first moment seems to be awkward, but the philosophy does not differ much from constructing a triangulation for every loop like in \cite{ThiemannQSD1, ThiemannQSD4}. As argued above, this is necessary in order to obtain a non-graph changing, well-defined and cylindrically consistent operator. On a lattice these problems can be avoided. However, the lattice approach has several other disadvantages.

Together with \eqref{regHAm}, this furnishes the regularisation of $\widehat{\PR}$. Since it is possible to choose the second regularisation parameter $\delta$ sufficiently small such that 
\bq
[h_{e_l(\Delta)}^{-\epsilon_L}, \sqrt{W_r(\delta, v(\Delta))}]
\eq
vanishes if $v(\Delta)$ does not lie on $\gamma$, the operator regulated as above acts on on a spin net function $f_{\gamma_r}$ as:
\begin{align}
\label{finalaction}
\begin{split}
&\widehat{{H}_{\cal T}^{r}} f_{\gamma_r}
=\frac{1}{\|\gamma^{(0)}_{prop}\|\,\beta^2  \, C(k)} \sum_{v\in\gamma^{(0)}_{prop}}
\left(\frac{n}{E(v)}\right)^2 \sum_{v(\Delta),v(\Delta')=v}
\epsilon^{IJ}\epsilon^{KL} \\
& ~~~~~~~~~ ~~~~\times\tr\left[ h_{\alpha_{IJ}(\Delta)} h^{\epsilon_K}_{e_K(\Delta')} \left[h^{-\epsilon_K}_{e_K(\Delta')}, \sqrt{W_r(\delta, v)}\right] 
 h^{\epsilon_L}_{e_L(\Delta')}\left[h^{-\epsilon_L}_{e_L(\Delta')}, \sqrt{W_r(\delta, v)}\right] \right]f_{\gamma_r} ~. 
 \end{split}
 \end{align}
Note that the right hand side of \eqref{finalaction} does not depend on the regulator $\delta$ as long as it small enough such that $W_r(\delta, v)$ only acts on the vertex $v$. Therefore we will drop $\delta$ in the following. At first sight, it seems, however, that the dependence on the triangulation can not be removed since the triangulation is bound to the minimal loops in $\gamma$. Yet, once spherical diffeomorphism invariance is implemented, the size of the loops becomes meaningless. This implies that the limit $\eta\to0$ can be taken in the weak ${\cal D}^{\ast}_{\text{diff}}$ topology in the same manner as this limit is taken in \cite{ThiemannQSD1}. Therefore, we will denote the `Hamiltonian' just by $\widehat{H_r}$ in the following. One should, nevertheless, be aware that in the above context implementing spherical diffeomorphisms corresponds to restricting to spherical symmetry as the system is already gauge fixed classically. 

Before we move on, let us quickly discuss why we do not regularise $\PR$ on a fixed (generalised) triangulation of which $\gamma$ is the one-skeleton, which seems to be more natural. The problem is that if one would consider such a regularisation, the operator $\widehat{\PR}$ should also see the edges of $\gamma$ labelled by the trivial representation, which are needed to complete the triangulation. But this implies that the holonomy $h_{\alpha}$ associated to the curvature can change the trivial colouring of an edge to a non-trivial one, so that the resulting operator is not graph preserving in a strict sense. In particular, a kink state would no-longer be described by a single $j$ once $\PR$ has acted and, thus, result in a more complicate comparison to the mini-superspace model. Furthermore, such an operator would no longer be cylindrically consistent in the usual sense, since it would depend on the chosen extension of $\gamma$ to a one-skeleton of a triangulation. Lastly, in this case one would need to construct a continuum limit, e.g. by coarse graining, which goes beyond the scope of this paper.

\section{Properties of $\widehat{\PR^r}$ and action on a kink}

The operator $\widehat{{H}_{r}}$ constructed in the previous section is trivially cylindrically consistent and so is 
\bq
\widehat{\PR^r}=i\frac{\sqrt{4\pi}}{\beta^2 } \left[\widehat{{H}_{r}},\sqrt{\widehat{W_r}} \right]
\eq
because neither $\widehat{{H}_{r}}$ nor $\widehat{W_r}$ act on edges labelled by the trivial representation, nor on 2-valent vertices with colinear tangents. By construction, both operators, $\widehat{{H}_{r}}$ and $\widehat{W_r}$, are also diffeomorphism covariant and have a non-trivial action. Thus, they fulfil all requirements for being well-defined in the full space and can be projected to the quantum reduced space (spherically diffeomorphism invariant space). Albeit, $\widehat{{H}_{r}}$ is not symmetric and can also not be extended to a symmetric operator for the following reason: Suppose $\widehat{{H}_{r}}$ acts on a vertex adjacent to an edge coloured by the same representation $k$ in which the holonomies of $\widehat{{H}_{r}}$ are regulated, then, by coupling the loop $h_{\alpha}$, the representation on the edge can change to the trivial representation. In this sense, $\widehat{{H}_{r}}$ can delete an edge. Thus, the dual of $\widehat{{H}_{r}}$ must create an edge. But $\widehat{{H}_{r}}$ is constructed in such a way that it can not create an edge which implies that $\widehat{{H}_{r}}$ can not coincide with his dual. Consequently, $\widehat{\PR^r}$ is not symmetric in contrast to the operator one obtains when quantising the classically reduced system. To circumvent this issue, one has to chose a different strategy to regularise $\widehat{{H}_{r}}$, for example the graph-changing regularisation proposed in \cite{LewandowskiSymmetricScalarConstraint} or work on a fixed lattice. 

\begin{figure}[t]
\center{\includegraphics{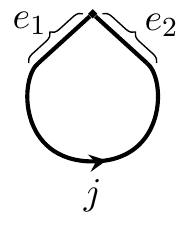}}
\captionsetup{width=0.6\textwidth}
\caption{\label{fig:loop}An anticlockwise oriented loop with a kink labelled by $j$. $e_1$ and $e_2$ are the 'sides' of the polyhedron connected by a `loop'  in the regularisation of $\PR$.}
\end{figure}

In general, $\widehat{{H}_r}$, and thus $\widehat{\PR}$, act by raising and lowering the spin on the minimal loops. Their action can be evaluated via the graphical calculus \cite{AlesciLinkingCovariantAnd, AlesciMatrixElementsOf}. On a loop with a kink (Fig.\ref{fig:loop}) embedded into the sphere $S^2_r$, the `Hamiltonian' acts as follows (for more details see appendix \ref{app:evaluation}):
\begin{align}
\label{action_on_kink}
\begin{split}
&\Hop \ket{j}_{\text{kink}}
= 
\frac{n^2}{\beta^2 C(k)} \epsilon^{IJ}\epsilon^{KL} 
\tr\left(h_{\alpha_{IJ}} h^{\epsilon_K}_{e_K} \left[h^{-\epsilon_K}_{e_K},\sqrt{\widehat{W_r}}\right] h^{\epsilon_L}_{e_L}\left[h^{-\epsilon_L}_{e_L},\sqrt{\widehat{W_r}}\right] \right) \ket{j}_{\text{kink}}
\\
&= \frac{2 n^2}{\beta^2 C(k)} \sum_{\tilde{j}_1=|j-k|}^{j+k}d_{\tilde{j}_1}\sum_{\tilde{j}_2=|j-k|}^{j+k}d_{\tilde{j}_2} \sqrt{W(j,\tilde{j}_1,k)} (-)^{k+j+\tilde{j}_2}
\sum_{\tilde{k}=0}^{2k} d_{\tilde{k}} (1-(-)^{\tilde{k}}) 
\sixj{k}{j}{j}{k}{\tilde{j}_2}{\tilde{k}}
\\
&\times
\left(\frac{\delta_{\tilde{j}_1,\tilde{j}_2}}{d_{\tilde{j}}}\left[\sqrt{W(j,j,0)} - \sqrt{W(j,j,\tilde{k})}\right]-(-)^{2\tilde{j}_1}\sqrt{W(j,\tilde{j}_2,k)} \sixj{\tilde{j}_1}{\tilde{j}_2}{j}{j}{k}{k}\right)
\\
&\times
\sum_{\tilde{j}_3=|j-k|}^{j+k} (-)^{j+\tilde{j}_3+k} \sixj{j}{k}{k}{j}{\tilde{j}_3}{\tilde{k}} \ket{\tilde{j}_3}_{\text{kink}}
\end{split}
\end{align}
Here, $d_j$ is the dimension of the irrep $j$, i.e. $d_j=2j+1$, and $W(j,\tilde{j},k)$ is the eigenvalue of $\widehat{W_r}$ when acting on a 2-valent gauge variant\footnote{The action of $\widehat{W}_r$ on an n-valent gauge variant vertex transforming in $k$ is essentially given by the action on an $n+1$-valent gauge invariant vertex, just that $\widehat{W_r}$ does not act on the artificial leg labelled by $k$.}  node transforming in the representation $k$ whose adjacent edges are labelled by $j$ and $\tilde{j}$. The action of $\widehat{W_r}$ can in principle also be evaluated using graphical calculus. Yet, the evaluation by algebraical means is straight forward on a 2-valent vertex (see \cite{ThiemannQSD4} for details). It yields
\bq
W(j,\tilde{j},k)= \frac{\beta^2}{4} \sqrt{\Delta_j\Delta_{\tilde{j}}-\frac{1}{4}(\Delta_k-\Delta_j -\Delta_{\tilde{j}})^2+\frac{1}{2}(\Delta_k-\Delta_j -\Delta_{\tilde{j}})}
\eq
where $\Delta_j=-j(j+1)$ which reduces to $\frac{\beta^2}{4}\sqrt{j(j+1)}$ on a gauge invariant node (i.e. if $k=0$ and $j=\tilde{j}$). Thus acting with $\PR$ on a loop with a kink gives
\be
\label{eq:1.reg}
\widehat{\PR} \ket{j}= i \frac{\sqrt{\pi}}{\beta} \sum_{\tilde{j}=|j-k|}^{j+k} \bra{\tilde{j}}\widehat{{H}_r}\ket{ j}((j(j+1))^{1/4} - (\tilde{j}(\tilde{j}+1))^{1/4})
 \ket{\tilde{j}}
\ee
while $\widehat{\R}_r$ simply yields 
\be
\label{eq:R full}
\widehat{\R_r} \ket{j}_{\text{kink}} =  \frac{1}{\sqrt{4\pi}} \sqrt{W(j,j,0)}\ket{j}_{\text{kink}} = \frac{ \beta}{2 \sqrt{4\pi}} (j(j+1))^{1/4} \ket{j}_{\text{kink}}
\ee
If we regularise $\widehat{\PR}$ in the defining representation $1/2$ or in spin $1$, then ${\Hop}$ simplifies to
\begin{align}
\begin{split}
&\Hop \ket{j}_{\text{kink}}
= \frac{12\, n^2}{\beta^2 C(1/2)} \sum_{\tilde{j}_1=j-k}^{j+k}d_{\tilde{j}_1}\sum_{\tilde{j}_2=|j-k|}^{j+k}d_{\tilde{j}_2} \sqrt{W(j,\tilde{j}_1,k)} (-)^{k+j+\tilde{j}_2}
\sixj{k}{j}{j}{k}{\tilde{j}_2}{1}
\\
&\times
\left(\frac{\delta_{\tilde{j}_1,\tilde{j}_2}}{d_{\tilde{j}}}\left[\sqrt{W(j,j,0)} - \sqrt{W(j,j,1)}\right]-(-)^{2\tilde{j}_1}\sqrt{W(j,\tilde{j}_2,k)} \sixj{\tilde{j}_1}{\tilde{j}_2}{j}{j}{k}{k}\right)
\\
&\times
\sum_{\tilde{j}_3=j-k}^{j+k} (-)^{j+\tilde{j}_3+k} \sixj{j}{k}{k}{j}{\tilde{j}_3}{1} \ket{\tilde{j}_3}_{\text{kink}}
\\
&= \frac{27}{\beta^2 C(k)} \sum_{\tilde{j}_1=j-k}^{j+k}d_{\tilde{j}_1}\sum_{\tilde{j}_2=|j-k|}^{j+k}d_{\tilde{j}_2} \sqrt{W(j,\tilde{j}_1,k)} \,
\frac{\Delta_{\tilde{j}_2}-\Delta_j-\Delta_k}{\sqrt{d_j\Delta_j d_k \Delta_k}}
 \\
&\times
\left(\frac{\delta_{\tilde{j}_1,\tilde{j}_2}}{d_{\tilde{j}}} \frac{\beta}{2}\left[(-\Delta_j)^{1/4} - (-2-3\Delta_j)^{1/4}\right]-(-)^{2\tilde{j}_1}\sqrt{W(j,\tilde{j}_2,k)} \sixj{\tilde{j}_1}{\tilde{j}_2}{j}{j}{k}{k}\right)
\\
&\times
\sum_{\tilde{j}_3=j-k}^{j+k} \frac{\Delta_{\tilde{j}_3}-\Delta_j-\Delta_k}{\sqrt{d_j\Delta_jd_k \Delta_k}}
\ket{\tilde{j}_3}_{\text{kink}}
\end{split}
\end{align}
where in the last step we used 
\bq
(-)^{k+j+\tilde{j}_2}
\sixj{k}{j}{j}{k}{\tilde{j}}{1}=\frac{\Delta_{\tilde{j}}-\Delta_j-\Delta_k}{2\sqrt{ d_j\Delta_j d_k\Delta_k}}~.
\eq
One would hope that $\widehat{\PR}$ and $\widehat{\R}$ reproduce the classical algebra of the reduced theory up to quantum corrections which should vanish if $j$ becomes large. However, for $k=1/2$ we find by Taylor expanding in $1/j$
\be
\label{eq:P1}
\bra{j\pm1/2}\widehat{\PR}\ket{j} \approx  \frac{ i \sqrt{4\pi}}{\beta}\, 0.04 \,n^2 \, j^{-1/2}+ {\cal O}(j^{-3/2})
\ee
and 
\be
\label{eq:P1_R}
\bra{j\pm1/2}[\widehat{\PR},\widehat{R}]\ket{j}\approx\mp i\,0.005\,n^2\,j^{-1} +{\cal O}(j^{-2}) \text{,}
\ee
where the coefficients $0.04$ and $0.005$ are numerical approximations of the analytic coefficients and $n=3$ or $n=4$ depending on the regularisation. The diagonal matrix elements vanish\footnote{This is of course the case in any regularisation $k$}. In the limit $j\to\infty$, the operators $\PR$ and $\R$ therefore commute on 2-valent vertices, which is counter intuitive. One might wonder whether this behaviour changes when regularising $\PR$ in a different spin. Yet, also for $k=1$ one finds a very similar behaviour where only the numerical coefficients  slightly differ. As the dominant contribution in the large spin limit steams from the differences in $W_r$  and not from the $6j$-symbols, there is little chance to change this behaviour by considering even higher regularising spins. Nevertheless, the scaling behaviour can be changed by modifying the regularisation of $\PR$ as will be discussed now.

Classically, $\{{H},V({\cal R})\}$ is equivalent to $\epsilon^{AB} n^i \{F^i_{AB}, V({\cal R})\}$ since $n^i$ obviously Poisson commutes with the volume. Starting with the latter expression and performing all the regularisation steps as described above one obtains the operator 
\begin{align} \nonumber
&\widehat{\PR^{(2)}}f_{\gamma_r}
=\frac{i\sqrt{4\pi}}{\|\gamma^{(0)}_{prop}\|\,\beta^4  \, C(k)} \sum_{v\in\gamma^{(0)}_{prop}}
\left(\frac{n}{E(v)}\right)^2 \sum_{v(\Delta),v(\Delta')=v}
\epsilon^{IJ}\epsilon^{KL} \\
&\times\tr\left( \left[h_{\alpha_{IJ}(\Delta)},\sqrt{\widehat{W_r}}\right]h^{\epsilon_K}_{e_K(\Delta')} \left[h^{-\epsilon_K}_{e_K(\Delta')}, \sqrt{W_r(\delta, v)}\right] 
 h^{\epsilon_L}_{e_L(\Delta')}\left[h^{-\epsilon_L}_{e_L(\Delta')}, \sqrt{W_r(\delta, v)}\right] \right)f_{\gamma_r} 
\end{align}   
that acts on a kink by 
\begin{align} \nonumber
&\widehat{\PR^{(2)}}\ket{j}_{\text{kink}}
= \frac{i 2 n^2\sqrt{4\pi}}{\beta^4 C(k)}
 \sum_{\tilde{j}_1=|j-k|}^{j+k}d_{\tilde{j}_1}\sum_{\tilde{j}_2=|j-k|}^{j+k}d_{\tilde{j}_2} \sqrt{W(j,\tilde{j}_1,k)} (-)^{k+j+\tilde{j}_2}
\sum_{\tilde{k}=0}^{2k} d_{\tilde{k}} (1-(-)^{\tilde{k}}) 
\sixj{k}{j}{j}{k}{\tilde{j}_2}{\tilde{k}}
\\
&\times \nonumber
\left(\frac{\delta_{\tilde{j}_1,\tilde{j}_2}}{d_{\tilde{j}}}\left[\sqrt{W(j,j,0)} - \sqrt{W(j,j,\tilde{k})}\right]-(-)^{2\tilde{j}_1}\sqrt{W(j,\tilde{j}_2,k)} \sixj{\tilde{j}_1}{\tilde{j}_2}{j}{j}{k}{k}\right)
\\
&\times 
\sum_{\tilde{j}_3=|j-k|}^{j+k} (-)^{j+\tilde{j}_3+k} \sixj{j}{k}{k}{j}{\tilde{j}_3}{\tilde{k}} 
\left(\sqrt{W(j,j,\tilde{k})}-\sqrt{W(\tilde{j}_3,\tilde{j}_3,0)}
\right)
\ket{\tilde{j}_3}_{\text{kink}}~.
\end{align}
The only difference to the earlier version is the term $\left(\sqrt{W(j,j,\tilde{k})}-\sqrt{W(\tilde{j}_3,\tilde{j}_3,0)}
\right)$ instead of $\left(\sqrt{W(j,j,0)}-\sqrt{W(\tilde{j}_3,\tilde{j}_3,0)}
\right)$, which one obtains from the commutator of $\sqrt{W}$ with $h_{\alpha}$ respectively $H$. What on the first sight appears like an ordering ambiguity has a tremendous influence on the scaling behaviour. This is because the leading order of the first bracket is $j^{1/2}$ while the leading order of the second one is $j^{-1/2}$. For $k=1/2$ one now obtains  
\be
\bra{j\pm1/2}\widehat{\PR^{(2)}}\ket{j} \approx\mp 0.06 n^2  \frac{ i \sqrt{4\pi}}{\beta}\, j^{1/2}+ {\cal O}(j^{-1/2})
\ee
and 
\be
\bra{j\pm1/2}[\widehat{\PR^{(2)}},\widehat{R}]\ket{j}\approx -i\,0.007\,n^2+{\cal O}(j^{-1})~. \label{eq:QuantumAlgebraGoodRegularisation}
\ee
Thus $\widehat{\PR^{(2)}}$ and $\widehat{R}$ do not commute in the large $j$ limit but still don't reproduce the classical algebra quantitatively, as we will see in the next section. Despite of these results one should be careful with concluding that the second regularisation is preferred over the first as the described change in the scaling behaviour might very well be an artefact of using 2-valent vertices. Already on a 3-valent node the action of $\PR$ is much more complicated since $\hat{W}$ is no longer diagonal on the gauge-variant nodes appearing during the evaluation of $\PR$. Moreover, also the scaling behaviour of $\widehat{W}$ and thus of $\widehat{R}$ changes. To see this, consider a three-valent node where all edges carry the same representation $j$. According to \cite{ThiemannQSD4}, the eigenvalues of $\widehat{W}$ on a 3-valent gauge-invariant node are given by (see equation A.6 in \cite{ThiemannQSD4}):
\bq
W_3(j_1,j_2,j_3)=\frac{1}{4}\sqrt{\frac{9}{2}\left(\Delta_{j_1}\Delta_{j_2}+\Delta_{j_1} \Delta_{j_3}+\Delta_{j_2} \Delta_{j_3}\right)-\frac{9}{4}\left(\Delta_{j_1}^2+\Delta_{j_2}^2+\Delta_{j_3} ^2\right)-\frac{1}{2}\left(\Delta_{j_1}+\Delta_{j_2}+\Delta_{j_3} \right)}~.
\eq   
Thus if all spins are equal then the leading order of $W_3(j,j,j)$ is given by $\frac{1}{4}\frac{3\sqrt{3}}{2} j^2 $ and not by $\frac{1}{4}j$ as in the case of a 2-valent vertex. On the other hand, also the scaling behaviour of $\PR$ in the first regularisation \eqref{eq:1.reg} should change since now $\sqrt{W_3(j,j,j)}- \sqrt{W_3(\tilde{j}_1,\tilde{j_2},j)}$ scales as $c+{\cal O}(j^{-1})$ for some $c\neq0$ where the second 'volume' is evaluated after the action with $H_r$. A similar effect should also be expected in the case of the second regularisation. Of course, in order to determine the scaling behaviour of $\PR$ on a 3-valent vertex exactly, more work is needed.

\section{Comparison of the classically and quantum reduced algebras} \label{sec:Comparison}

Let us now compare the relation \eqref{eq:QuantumAlgebraGoodRegularisation} to its analogue in the reduced quantisation. Here, we a priori also don't have an operator corresponding $\tilde P_R$, but we can approximate it via the standard polymerisation 
\be 
	\tilde P_R \approx \sin (\mu \tilde P_R) / \mu = \frac{e^{i \mu \tilde P_R} - e^{-i \mu \tilde P_R}}{2 i \mu} \text{.}
\ee
Using this polymerisation and the short notation $\ket{\rho} = \exp \left(- i \, \rho \, \tilde P_\Lambda(r) \right)$, we find
\be
	\braopket{\rho}{\left[ \hat {\tilde P}_R(r), \hat {\tilde R}(e^r)\right]}{\rho'} = \frac{-i}{2} \left(\delta_{\rho,\rho'+\mu}+\delta_{\rho,\rho'-\mu} \right) \text{.} \label{eq:ReducedCommutator}
\ee
In order to match \eqref{eq:ReducedCommutator} to \eqref{eq:QuantumAlgebraGoodRegularisation} based on the identification $j = \rho$ for large quantum numbers as suggested in section \ref{sec:FullSpectralProperties}, one would now like to set $\mu = 1/2$. The qualitative structure of \eqref{eq:ReducedCommutator} and \eqref{eq:QuantumAlgebraGoodRegularisation} would then be the same, however the overall coefficient is off by a factor of about $72/n^2$, where $n$ equals $3$ or $4$ depending on the chosen regularisation. If we instead consider the first ordering of $\PR$, which is closer to \cite{ThiemannQSD4}, then \eqref{eq:ReducedCommutator} and \eqref{eq:P1_R} also show different scaling behaviours in the regime of large quantum numbers. Moreover, comparing the action of the fundamental operators $\PR$ and $\R$ in the mini-superspace model with the one in the full theory, we find  that in all considered regularisation schemes the scaling behaviour of the fundamental operators differ. That is, the leading order in $\R$ scales with $j^{1/2}$ (see \eqref{eq:R full}) in the full theory, while it is linear in $\rho$ in the mini-superspace model and $\PR$ scales with $j^{\pm 1/2}$ (see \eqref{eq:P1} and \eqref{eq:QuantumAlgebraGoodRegularisation}) in the full theory, while the leading order in the mini-superspace model is constant.

Does this show that it is not possible to recover the mini-super space model as a sub sector of the `finer' quantum reduced model? The answer to this question is clearly no. In contrast, the fact that it is possible to find a quantisation of $\PR$ and $\R$ in the full theory such that the commutators  \eqref{eq:ReducedCommutator} and \eqref{eq:QuantumAlgebraGoodRegularisation} qualitatively match is very promising. Moreover, there are several parameters that can be tuned to minimise the numerical mismatch, e.g. the regularising spin $k$ or the choice of triangulation. If, for example, one considers the mirror construction of \cite{ThiemannQSD4} with $n=4$ to complete the triangulation then \eqref{eq:ReducedCommutator} and \eqref{eq:QuantumAlgebraGoodRegularisation} only differ by roughly a factor of $4$. The strange scaling behaviours of $\PR$ and $\R$ in the full theory are an artefact of using kink states only. As argued at the end of the previous section, the scaling of the volume $W$ is different on 2-valent nodes since part of the contributions cancel, which is no longer true on higher-valent nodes.

The above results, thus, suggest that the mini-superspace model cannot be recovered in the kink sector of the full theory alone, but should be searched for by considering also higher-valent vertices. Yet, more general states in the quantum reduced theory are no longer labeled by a single spin, which makes it impossible to directly identify the quantum numbers in both approaches. Instead, one should expect that the classically reduced model arises a sort of coarse grained version of the quantum reduced model. This could look as follows: Instead of computing the reduced algebra $\PR,\R$ on kink states, one computes the algebra on the one-skeleton of a triangulation of the sphere and then construct the continuum limit by following e.g. the coarse graining program advertised in \cite{DittrichTheContinuumLimit}. As argued in \cite{DittrichHowToConstruct}, the continuum limit should be reached if (spherical) diffeomorphism invariance is reinstalled, i.e. in our context if spherical reduction is reached. Such a procedure also has the advantage that a regularisation of $\PR$ on a lattice is much more natural. To conclude,  the so far observed qualitative match of \eqref{eq:ReducedCommutator} and \eqref{eq:QuantumAlgebraGoodRegularisation} is promising, but a further refinement of the calculation is needed, which, however, goes beyond the scope of this paper.

\section{Conclusion} \label{sec:Conclusion}

In this paper, we have compared the main algebraic relation of general relativity classically reduced to spherical symmetry to its counterpart in a theory reduced at the quantum level. We have found a certain qualitative match, the precise numerical prefactor is however clearly off. We have attributed this issue to the special ``degenerate'' nature of the quantum states used. Given more complicated quantum reduced full theory states corresponding to proper triangulations of the spheres $S^2_r$, we argued that certain insufficiencies in the calculations would be improved upon. However, in this case, the identification of full theory and reduced theory quantum states is less straight forward as it will involve a non-trivial form of coarse graining. This is in particular important for a  physically viable definition of the dynamics, which, in the reduced theory, might also need a modification via a $\bar \mu$-type \cite{AshtekarQuantumNatureOf} prescription as in the cosmological case. In the full theory, due to the gauge group SU$(2)$, such a modification cannot be implemented for the kink-states used in this paper, whereas it can in the reduced quantisation where the $\mathbb R_\text{Bohr}$ allows for arbitrarily small quantum numbers. A $\bar \mu$-prescription would thus have to arise from coarse graining the fundamental SU$(2)$-based dynamics. The simplified setting of evaluating the commutator $\left[ \hat {\tilde P}_R, \hat {\tilde R}\right]$ seems ideal for attacking this type of question. 

It is also interesting to ask how our work compares to other approaches to spherical symmetry in the literature. In the framework based on group field theory condensates \cite{GielenCosmologyFromGroup, OritiGeneralizedQuantumGravity}, coarse graining seems to be implemented at a more fundamental level. It would be interesting to investigate such condensates, e.g. a condensate of kink states, also within our framework. 
Also, it would be interesting to approach the question of defining quantum states corresponding to black holes, following \cite{GambiniLoopQuantizationOf}. 
In view of the different spectra of geometric operators that can be obtained (see section \ref{sec:ReducedSpectralProperties}), care should be taken whenever arguments based on such spectra are made. First and foremost however, the question of gravitational collapse \cite{JoshiRecentDevelopmentsIn} seems most pressing at the moment, see e.g. \cite{AshtekarBlackHoleEvaporation, BojowaldLemaitreTolmanBondi, GambiniQuantumShellsIn} for some seminal work in the context of loop quantum gravity. Here, the fate of the singularity in the centre of classical black holes and possible observational consequences resulting from its resolution due to quantum gravity effects \cite{RovelliPlanckStars, HaggardBlackHoleFireworks, BarrauFastRadioBursts} are of key importance.

\section*{Acknowledgements}
NB was supported by a Feodor Lynen Research Fellowship of the Alexander von Humboldt-Foundation during parts of this work. AZ and NB (partially) were supported by the Polish National Science Centre grant No. 2012/05/E/ST2/03308. 
Discussions with Jerzy Lewandowski and J{\k{e}}drek \'Swie\.zewski are gratefully acknowledged.

\appendix

\section{Evaluation of $\Hop$} 
\label{app:evaluation}
In order to evaluate the action of $\Hop$, we use the graphical calculus of \cite{AlesciLinkingCovariantAnd, AlesciMatrixElementsOf}. In this calculus, the Wigner matrices are represented by a line with a dart symbolising the group element, i.e. 
\bq
\vspace*{-5pt}
[R^{j}(g)]^{m}_{\;\;\;n}=\makeSymbol{\raisebox{.4\height}{\includegraphics[scale=0.8]{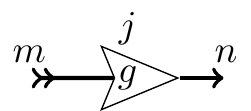}}}
\eq 
where $m,n$ are magnetic numbers. 3 j-symbols are depicted by 3-valent nodes, that is 
\bq
\makeSymbol{\includegraphics[scale=0.8]{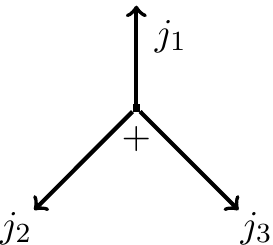}}=\threej{j_1}{m_1}{j_2}{m_2}{j_3}{m_3}
\eq
where $+$ indicates anticlockwise ordering and $-$ clockwise ordering. Inner legs of the intertwiner not representing a true edge of the graph are drawn as dashed lines. To evaluate the action of $\Hop$, we essentially only need three identities:
\begin{itemize}
\item[A)] Coupling identity
\vspace*{-5pt}
\be
\label{coupling}
\makeSymbol{\includegraphics[scale=0.8]{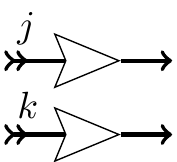}}\;\;=\sum_{\tilde{j}=|j-k|}^{j+k} d_{\tilde{j}} \makeSymbol{\includegraphics[scale=0.8]{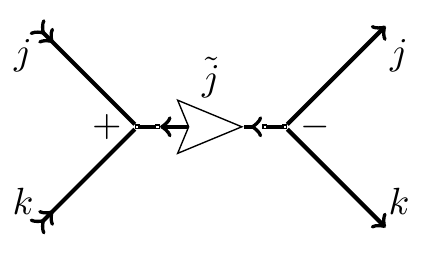}}=\sum_{\tilde{j}=|j-k|}^{j+k} d_{\tilde{j}} (-)^{2\tilde{j}}\makeSymbol{\includegraphics[scale=0.8]{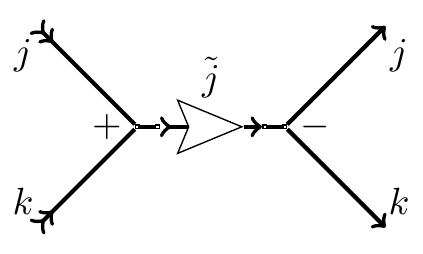}}
\ee
The middle leg in the second equality carries the adjoint matrix element $\overline{R(g)}$, which is symbolised by the inner arrows pointing against the direction of the dart. Using the properties of Wigner matrices one can show that this is equivalent to replacing $\overline{R^m\,_{n} (g)}$ by $R^n\,_{m} (g)\,(-)^{2j}$ when contracting $m,n$ with the corresponding $3j$ symbols (last equality). For simplicity the darts representing the group elements will be left away in the following.  

\item[B)] Orthogonality relation\vspace*{-5pt}
\be
\label{ortho}
\makeSymbol{\includegraphics[scale=0.8]{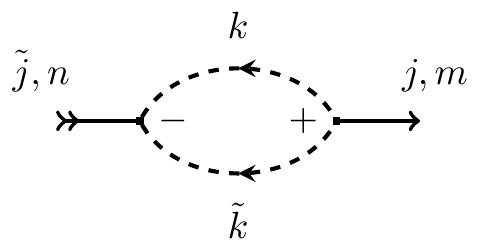}}=\frac{1}{d_c}
\makeSymbol{\raisebox{.75\height}{\includegraphics[scale=0.8]{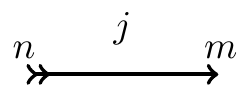}}}
\ee
\vspace*{-5pt}
\item[C)] $6j$ identity
\vspace*{-5pt}
\be
\label{6j}
\makeSymbol{\includegraphics[scale=0.8]{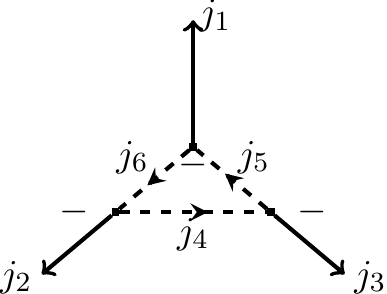}}
 =\sixj{j_1}{j_4}{j_2}{j_5}{j_3}{j_6} \makeSymbol{\includegraphics[scale=0.8]{kink-figure4.pdf}}
\ee
\vspace*{-5pt}
\end{itemize}
Furthermore, one should keep in mind that changing the orientation of an edge labelled by $j$ produces a sign $(-)^{2j}$ and changing the ordering of a node with adjacent legs $j_1$,$j_2$ and $j_3$ yields a sign $(-)^{j_1+j_2+j_3}$. Moreover, if $j_1$, $j_2$ and $j_3$ are coupled then by the triangular identity $(-)^{2(j_1+j_2+j_3)}=1$ holds.

In the subsequent, we want to evaluate the action of $\Hop$ on a counterclockwise oriented loop with a kink  labelled by $j$ (see Fig.\ref{fig:loop}). On such a state, the action of $\Hop$ reduces to:
\begin{align}
\Hop \ket{j}
&= \frac{9}{\beta^2 C(k)}\tr\left[(h_{\alpha}-h_{\alpha}^{-1})\vphantom{ \left[\sqrt{\widehat{W_r}},h_{s_1}\sqrt{\widehat{W_r}} h_{s_1}^{-1}\right]}\right.\\
&
\times
\left. \left(\left[\sqrt{\widehat{W_r}},h_{e_1}\sqrt{\widehat{W_r}} h_{e_1}^{-1}\right]
-\left[\sqrt{\widehat{W_r}},h^{-1}_{e_2}\sqrt{\widehat{W_r}} h_{e_2}\right] +\left[h_{e_1}\sqrt{\widehat{W_r}}h_{e_1}^{-1},h^{-1}_{e_2}\sqrt{\widehat{W_r}} h_{e_2}\right]\right)\right]
 \ket{j} \nonumber
 \end{align}
where $\alpha$ is the counterclockwise oriented loop, $e_1$ is the outgoing half-edge at the kink and $e_2$ the ingoing one. 
 
Using \eqref{coupling} and the fact that $\widehat{W}$ is diagonal on 2-valent non-invariant nodes, we obtain:
\vspace*{-10pt}\mbox{}
\begin{align}
\label{eq:eval1}
\vspace*{-5pt}
\begin{split}
&
h_{e_1}\sqrt{\widehat{W_r}} h_{e_1}^{-1} \ket{j}
=h_{e_1}\sqrt{\widehat{W_r}}\sum_{\tilde{j}_1} d_{\tilde{j}_1} (-)^{2 k}\makeSymbol{\includegraphics[scale=0.8]{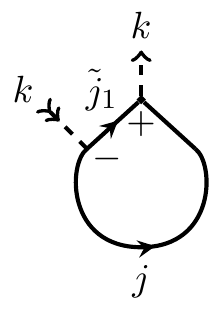}}
=\sum_{\tilde{j}_1} d_{\tilde{j}_1} (-)^{2 k} \sqrt{W(j,\tilde{j}, k)} \;\;\makeSymbol{\includegraphics[scale=0.8]{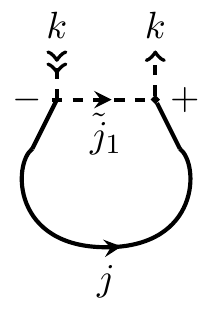}}\\[-15pt]
&=\sum_{\tilde{j}_1} d_{\tilde{j}_1} \sqrt{W(j,\tilde{j}, k)}\sum_{\tilde{k}} d_{\tilde{k}} \makeSymbol{\includegraphics[scale=0.8]{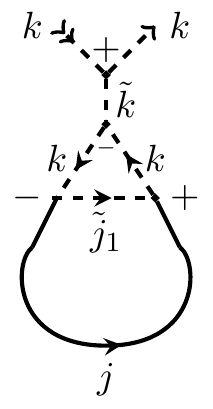}}\\[-15pt]
&=\sum_{\tilde{j}_1} d_{\tilde{j}_1} \sqrt{W(j,\tilde{j}, k)}\sum_{\tilde{k}} d_{\tilde{k}} (-)^{j+\tilde{j}_1+k} \sixj{j}{k}{k}{j}{\tilde{j}_1}{\tilde{k}} \makeSymbol{\includegraphics[scale=0.8]{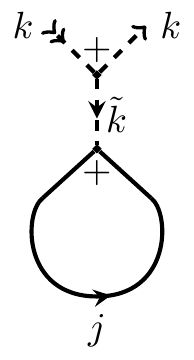}}
\end{split}
\end{align}
\vspace*{-20pt}\mbox{}
where in the last step we used \eqref{6j}. Similarly,  
\begin{align}
\vspace*{-5pt}
&h_{e_2}^{-1}\sqrt{\widehat{W_r}} h_{e_2} \ket{j}
=\sum_{\tilde{j}_1} d_{\tilde{j}_1} \sqrt{W(j,\tilde{j}, k)}\sum_{\tilde{k}} d_{\tilde{k}} (-)^{j+\tilde{j}_1+k+\tilde{k}} \sixj{j}{k}{k}{j}{\tilde{j}_1}{\tilde{k}} \makeSymbol{\includegraphics[scale=0.8]{kink-figure13.pdf}}
\\[-30pt]\mbox{} \nonumber
\end{align}
Thus the first two commutators yield:\\\mbox{}
\begin{align} 
& \left[\sqrt{\widehat{W_r}},h_{e_1}\sqrt{\widehat{W_r}} h_{e_1}^{-1}\right]
-\left[\sqrt{\widehat{W_r}},h^{-1}_{e_2}\sqrt{\widehat{W_r}} h_{e_2}\right]\ket{j}
= 
\\[-10pt]
&\sum_{\tilde{j}_1} d_{\tilde{j}_1} \sqrt{W(j,\tilde{j}, k)}\sum_{\tilde{k}} d_{\tilde{k}} (-)^{j+\tilde{j}_1+k}(1-(-)^{\tilde{k}}) \sixj{j}{k}{k}{j}{\tilde{j}_1}{\tilde{k}} \left(\sqrt{W(j,j,\tilde{k})}-\sqrt{W(j,j,0)}\right) \makeSymbol{\includegraphics[scale=0.8]{kink-figure13.pdf}} \nonumber
\end{align}\vspace*{-10pt}\mbox{} 
The third commutator is a bit harder to evaluate. It yields:
\begin{align}
&h^{-1}_{e_2}\sqrt{\widehat{W_r}} h_{e_2} h_{e_1}\sqrt{\widehat{W_r}} h_{e_1}^{-1} \ket{j} \nonumber
=h^{-1}_{e_2}\sqrt{\widehat{W_r}}h_{e_2} \sum_{\tilde{j}_1} d_{\tilde{j}_1} (-)^{2k} \sqrt{W(j,\tilde{j}_1,k)} \;\;\makeSymbol{\includegraphics[scale=0.8]{kink-figure11.pdf}}
\\[-15pt] \nonumber
&=h_{e_2}^{-1}\sqrt{\widehat{W_r}} \sum_{\tilde{j}_1,\tilde{j}_2} d_{\tilde{j}_1} d_{\tilde{j}_2} (-)^{2k} \sqrt{W(j,\tilde{j}_1,k)}\;\; \makeSymbol{\includegraphics[scale=0.8]{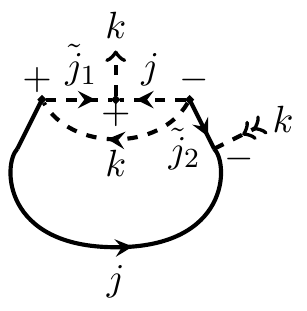}}
\\[-15pt]
&=
\sum_{\tilde{j}_1,\tilde{j}_2} d_{\tilde{j}_1} d_{\tilde{j}_2} \sqrt{W(j,\tilde{j}_1,k)} (-)^{2\tilde{j}_1} \sixj{j}{j}{k}{k}{\tilde{j_1}}{\tilde{j_2}}   \sqrt{W(j,\tilde{j}_2,k)}\;\; \makeSymbol{\includegraphics[scale=0.8]{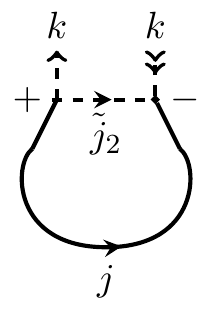}}
\\[-15pt] \nonumber
&=\sum_{\tilde{j}_1,\tilde{j}_2} d_{\tilde{j}_1} d_{\tilde{j}_2} \sqrt{W(j,\tilde{j}_1,k)} (-)^{2\tilde{j}_1} \sixj{j}{j}{k}{k}{\tilde{j_1}}{\tilde{j_2}}   \sqrt{W(j,\tilde{j}_2,k)}\sum_{\tilde{k}}d_{\tilde{k}}(-)^{k+\tilde{j}_2+j+\tilde{k}} \sixj{j}{k}{k}{j}{\tilde{j}_2}{\tilde{k}}\makeSymbol{\includegraphics[scale=0.8]{kink-figure13.pdf}}
\end{align}
From the third to the fourth step, we first used \eqref{6j} in order to remove the inner dashed leg and then applied $\hat{W}$ to the resulting non-invariant node. To derive the last equality, the same coupling strategy as in \eqref{eq:eval1} was employed. The second term of the commutator can be evaluated by similar means and agrees with the expression above except for a sign $(-)^{-\tilde{k}}$. In total, one obtains 
\begin{align} \nonumber
\left[h_{e_1}\sqrt{\widehat{W_r}}h_{e_1}^{-1},h^{-1}_{e_2}\sqrt{\widehat{W_r}} h_{e_2}\right]\ket{j}
&=\sum_{\tilde{j}_1,\tilde{j}_2} d_{\tilde{j}_1} d_{\tilde{j}_2} \sqrt{W(j,\tilde{j}_1,k)} (-)^{2\tilde{j}_1} \sixj{j}{j}{k}{k}{\tilde{j_1}}{\tilde{j_2}}   \sqrt{W(j,\tilde{j}_2,k)}
\\
&\times\sum_{\tilde{k}}d_{\tilde{k}}(-)^{k+\tilde{j}_2+j+\tilde{k}} \sixj{j}{k}{k}{j}{\tilde{j}_2}{\tilde{k}} (1-(-)^{\tilde{k}})\makeSymbol{\includegraphics[scale=0.8]{kink-figure13.pdf}}
\\[-25pt]\mbox{} \nonumber
\end{align}
The loop can now be easily coupled using the loop trick \cite{AlesciMatrixElementsOf}, i.e.
\bq
h_{\alpha}-h_{\alpha}^{-1}=\sum_{\tilde{k}} d_{\tilde{k}} (1-(-)^{\tilde{k}})(-)^{ 2k}\makeSymbol{\includegraphics[scale=0.8]{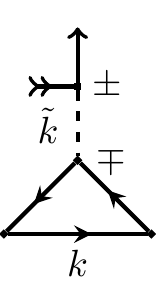}}~,
\\[-15pt]\mbox{}
\eq
which yields\\[-25pt]\mbox{}
\begin{align}
&\tr\big((h_{\alpha}-h^{-1}_{\alpha})\makeSymbol{\includegraphics[scale=0.8]{kink-figure13.pdf}}\big) \nonumber
=\sum_{\tilde{k}_2} d_{\tilde{k}_2} (1-(-)^{\tilde{k}_2})(-)^{2 k}\sum_{\tilde{j}} d_{\tilde{j}} (-)^{2\tilde{j}}\makeSymbol{\includegraphics[scale=0.8]{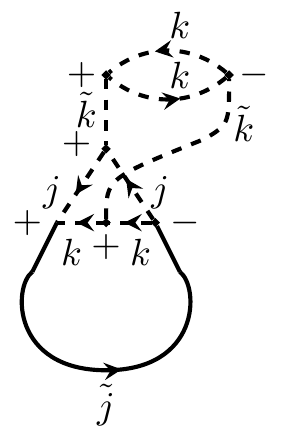}}
\\[-15pt] \nonumber
&=(1-(-)^{\tilde{k}})\sum_{\tilde{j}} d_{\tilde{j}} (-)^{2\tilde{j}}\;\;\makeSymbol{\includegraphics[scale=0.7]{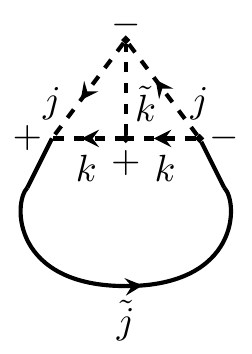}}
=(1-(-)^{\tilde{k}})\sum_{\tilde{j}} d_{\tilde{j}} (-)^{2\tilde{j}+\tilde{k}} \sixj{j}{k}{k}{j}{\tilde{j}}{\tilde{k}} \makeSymbol{\includegraphics[scale=0.7]{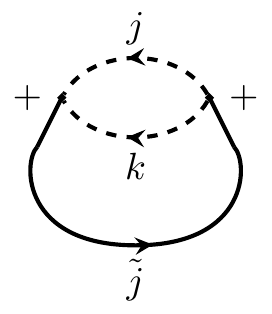}}
\\
&=((-)^{\tilde{k}}-1)\sum_{\tilde{j}} (-)^{j+\tilde{j}+k}\sixj{j}{k}{k}{j}{\tilde{j}}{\tilde{k}} \ket{\tilde{j}}
\end{align}
In the first and last equation, the orthogonality relation \eqref{ortho} was used, while in the second to last equation \eqref{6j} was exploited again.  
Adding up all contributions and using $(1-(-)^k)^2=2(1-(-)^k)$ finally leads to equation \eqref{action_on_kink}.


\end{document}